\documentclass[letter, 12pt]{article}

\usepackage[margin=1in]{geometry}
\usepackage[numbers,sort&compress]{natbib} 
\usepackage{authblk} 
\usepackage{color}
\usepackage{amsmath}
\usepackage{amssymb} 
\usepackage{graphicx} 
\usepackage{fancyhdr} 
\usepackage{amsfonts}
\usepackage{tabularx} 
\usepackage[
singlelinecheck=false 
]{caption}

\begin{document}

\rhead{OHSTPY-HEP-T-14-006}
\renewcommand\headrule{} 

\title{\huge \bf{Yukawa Unification in an SO(10) SUSY GUT: SUSY on the Edge}}
\author{Zijie Poh}
\author{Stuart Raby}
\affil{\emph{Department of Physics}\\\emph{The Ohio State University}\\\emph{191 W.~Woodruff Ave, Columbus, OH 43210, USA}}

\maketitle
\thispagestyle{fancy}
\pagenumbering{gobble} 

\begin{abstract}\normalsize\parindent 0pt\parskip 5pt
In this paper we analyze Yukawa unification in a three family SO(10) SUSY GUT.  We perform a global $\chi^2$ analysis
and show that SUSY effects do not decouple even though the universal scalar mass parameter at the GUT scale, $m_{16}$,  is found to lie
between 15 and 30 TeV with the best fit given for $m_{16} \approx 25$ TeV.   Note, SUSY effects don't decouple since stops and bottoms
have mass of order 5 TeV, due to RG running from $M_{GUT}$.   The model has many testable predictions.   Gauginos are the lightest sparticles and
the light Higgs boson is very much Standard Model-like.   The model is consistent with flavor and CP observables with the $BR(\mu \to e\gamma)$ close
to the experimental upper bound.   With such a large value of $m_{16}$ we clearly cannot be considered ``natural" SUSY nor are we ``Split" SUSY.  We are thus in the region in between or ``SUSY on the Edge."

\end{abstract}

\pagenumbering{arabic} 

\newpage
\section{Introduction}

Gauge and Yukawa unified $\text{SO}(10)$ supersymmetric grand unified theories (SUSY GUTs) \cite{Blazek:2001sb, Baer:2001yy, Blazek:2002ta, Tobe:2003bc, Auto:2003ys} are a class of highly constrained models.  By performing a global fit to low energy data, such as fermion masses and mixing angles and flavor observables, $\text{SO}(10)$ SUSY GUT models can provide experimental constraints on sparticle masses accessible at the LHC energy scale.  The specific $\text{SO}(10)$ SUSY GUT model that we study in this paper has a $D_3\times[\text{U}(1)\times\mathbb{Z}_3\times\mathbb{Z}_3]$ family symmetry \cite{Dermisek:2005ij, Dermisek:2006dc}.  This model includes three families of quarks and leptons and has been shown to provide good fits to low energy precision electroweak observables, including fermion masses and mixing angles \cite{Albrecht:2007ii, Anandakrishnan:2012tj, Anandakrishnan:2013cwa, Anandakrishnan:2013nca, Anandakrishnan:2014nea}.

In this paper we extend the analysis of previous papers by including additional low energy flavor observables.  We also identify which observables are better fit by incorporating the SUSY effects.   We then discuss the many experimental tests of this model. Finally we evaluate the amount of fine-tuning in this scenario.  As might be expected the amount of fine-tuning is enormous.  However, we show that introducing certain boundary conditions at the GUT scale has the effect of dramatically reducing the amount of fine-tuning.   The origin of such boundary conditions in this theoretical sweet spot is not known, although we discuss some possible sources in an Appendix.

The paper is organized as follows. In Section~\ref{sec:model}, we present an overview of the $\text{SO}(10)$ model and \texttt{maton}\footnote{``maton" was developed by Radovan Dermis\v{e}k and modified for large scalar masses by Archana Anandrakrishnan, B. Charles Bryant, Zijie Poh and Akin Wingerter.}, the program that runs all the parameters and calculates most of the observables.  In Section~\ref{sec:global_chi2}, we present the results of a global $\chi^2$ analysis for both the universal and ``mirage mediated" gaugino mass boundary conditions. We show that $\chi^2$ is minimized for values of the universal scalar mass, $m_{16}$ of order 25 TeV, for both universal gaugino masses at the GUT scale or ``mirage mediation" with non-universal gaugino masses, respectively (see Figs. \ref{fig:chi2_line_plot} and \ref{fig:contour}).   The $\chi^2$ function now includes the CP-even angular observables of $B^0\to K^*\mu^+\mu^-$ measured by LHCb.  Some of these observables are in slight tension with the standard model (SM) prediction \cite{Aaij:2013qta,Aaij:2013iag}.  In particular, the observable $P_5'$ integrated over the dimuon invariant mass squared, $1\leq q^2\leq6\ \text{GeV}^2$, has a $2.5\sigma$ discrepancy from the SM prediction.  In addition, the angular observable $P_4'$ integrated over $14.18\leq q^2\leq16\ \text{GeV}^2$ also has a $2.7\sigma$ tension with the SM prediction.  We also
compare our fits to inclusive vs. exclusive measurements of the CKM elements $V_{ub}$ and $V_{cb}$.

The first two families of squarks and sleptons have mass of order $m_{16}$ and thus they decouple from low energy physics.  However,  the third family scalars are significantly lighter, with mass between 4 and 10 TeV.   This is a consequence of a natural inverted scalar mass hierarchy \cite{Bagger:1999ty}.  Hence they do not decouple and thus contribute to lowering $\chi^2$.  In Section~\ref{ssec:b_physics} we discuss the observables which are better fit by including SUSY loops.  We then discuss bounds on the gluino mass in Section~\ref{ssec:gluino} and further predictions in Section~\ref{ssec:predictions}.

Since our scalar masses are heavy there is, in general, considerable fine-tuning of order 1 part in $10^5$.  In Section~\ref{sec:fine_tuning}, we present the results of a fine tuning analysis of the model and demonstrate that given particular ratios of soft breaking parameters at the GUT scale the amount of fine tuning is minimized to roughly 1 part in 500 (perhaps these boundaries conditions can be obtained in a fundamental SUSY breaking model).  Possible origins for the necessary boundary conditions are discussed in the Appendix.

Finally, we note that our soft breaking parameter $m_{16}$ lies in the range between 15 and 30 TeV.  Hence the gravitino mass is expected to be large, perhaps large enough to avoid the cosmological gravitino problem \cite{Weinberg:1982zq}.  The moduli may also be suitably heavy to avoid a cosmological moduli problem \cite{Coughlan:1983ci,Ellis:1986zt,Banks:1993en,deCarlos:1993jw}.  At the same time we see that SUSY effects have not decoupled from low energy flavor observables, thus our soft SUSY breaking parameters lie on the {\em edge} between {\em Split} and {\em natural} SUSY.  We denote this region of parameter space as {\em SUSY on the Edge}.  Gauginos are expected to be visible at the LHC with gluinos decaying predominantly into third generation quarks with special decay signatures.  While the light Higgs discovered by CMS and ATLAS is expected to be very much Standard Model-like.  We also present predictions for additional flavor and CP violating observables, see Table~\ref{tab:prediction}. In Section~\ref{sec:conclusion} we conclude and discuss phenomenological tests of our model.

\section{Model and Procedure}
\label{sec:model}

The details of our $\text{SO}(10)$ SUSY GUT model is presented in \cite{Anandakrishnan:2012tj}.  The GUT scale boundary conditions that we use are universal squark and slepton masses, $m_{16}$, universal trilinear couplings, $A_0$, and non-universal Higgs masses, $m_{H_u}$ and $m_{H_d}$.  As for the boundary condition for the gaugino masses, we consider two different cases.  One with universal gaugino masses, $M_{1/2}$, and another with mirage mediated gaugino masses.  The mirage mediated gaugino masses are defined as \cite{Anandakrishnan:2013cwa}
\begin{align}
  M_i = \left(1+\frac{g_G^2b_i\alpha}{16\pi^2}\log\frac{M_\text{pl}}{m_{16}}\right)M_{1/2} \,,  \label{eqn:mirage}
\end{align}
where $M_{1/2}$ and $\alpha$ are free parameters and $b_i=(33/5, 1, -3)$ for $i=1,2,3$ are the appropriate $\beta$-function coefficients.  Mirage boundary condition is interesting because $\alpha=1.5$ is the optimal scenario for a well-tempered dark matter \cite{Anandakrishnan:2014fia}.  On the other hand, it has been shown that the LSP of universal boundary condition is predominantly Bino-like \cite{Anandakrishnan:2013nca}, which leads to an over-closed universe.  However, this problem can be solved by introducing axions with mass lighter than the LSP into the model.  Hence, in this paper, we study the cases where the universal boundary condition, $\alpha=0$, and mirage boundary condition with $\alpha=1.5$.

On the other hand, the charged-fermion sector of our model has 12 parameters - 11 Yukawa parameters and $\tan\beta$.  In comparison, the SM has 13 free parameters.  Hence, our model has 1 prediction in the charged fermion sector.  Including the neutrino sector, our model has 3 additional free parameters to fit 6 observables - 2 $\Delta m^2$s, 3 real mixing angles and one CP violating phase.  Hence, our model has 4 predictions in the fermion sector.

To summarize, all the free parameters of our model are listed in Table \ref{tab:inputs}.  With universal gaugino masses boundary condition, our model has $24$ free parameters while with mirage mediated gaugino mass boundary conditions the number of free parameters increases to $25$.

\begin{table}
  \begin{center}
    \renewcommand{\arraystretch}{1.2}
      \scalebox{0.83}{
        \begin{tabular}{|l||c|c||c|c||}
          \hline Sector & Universal Gaugino Masses & No. & Mirage Mediated Gaugino Masses & No.
          \\ \hline Gauge & $\alpha_G$, $M_G$, $\epsilon_3$ & 3 & $\alpha_G$, $M_G$, $\epsilon_3$ & 3
          \\ SUSY (GUT scale) & $m_{16}$, $M_{1/2}$, $A_0$, $m_{H_u}$, $m_{H_d}$
            & 5 & $m_{16}$, $M_{1/2}$, $A_0$, $m_{H_u}$, $m_{H_d}$, $\alpha$ & 6
          \\ Yukawa Textures & $\epsilon$, $\epsilon'$, $\lambda$, $\rho$, $\sigma$, $\tilde\epsilon$, $\xi$, $\phi_{\rho}$, $\phi_{\sigma}$, $\phi_{\tilde\epsilon}$, $\phi_{\xi}$ & 11 & $\epsilon$, $\epsilon'$, $\lambda$, $\rho$, $\sigma$, $\tilde\epsilon$, $\xi$, $\phi_{\rho}$, $\phi_{\sigma}$, $\phi_{\tilde\epsilon}$, $\phi_{\xi}$ & 11
          \\ Neutrino & $M_{R_1}$, $M_{R_2}$, $M_{R_3}$ & 3 & $M_{R_1}$, $M_{R_2}$, $M_{R_3}$ & 3
          \\ SUSY (EW Scale) & $\tan\beta$, $\mu$ & 2 & $\tan\beta$, $\mu$ & 2
          \\ \hline Total & & 24 & & 25
          \\ \hline
        \end{tabular}
      }
    \caption{Our model has 24 free parameters for universal gaugino masses boundary conditions. For mirage mediated gaugino masses boundary condition, there is an additional input parameter, $\alpha$, which determines the amount of splitting of the gaugino masses at GUT scale.}
    \label{tab:inputs}
  \end{center}
\end{table}

The detailed procedure of our calculation is presented in \cite{Anandakrishnan:2012tj}.  In addition, we want to emphasize that after integrating out the right-handed neutrinos, we use the two-loop Minimal Supersymmetric Standard Model (MSSM) renormalization group equation (RGE) to run down to the weak scale.  At the weak scale, we include the one-loop threshold corrections for the Yukawa and the gauge couplings, which was calculated by Pierce et.~al.~\cite{Pierce:1996zz}.   The gluino mass, $M_{\tilde g}$, and CP odd Higgs mass, $M_A$, are pole masses.   However, we do not include one-loop threshold corrections for the other scalar masses.  Instead, we estimated that these corrections are about $10\%$.  Hence, the GUT Scale parameters of our model have an inherent $10\%$ theoretical error.  Adding the one-loop threshold correction for the soft scalar masses can be a future project.

In addition to calculating the observables included in \cite{Anandakrishnan:2012tj}, we included a low $q^2$ bin ($1\leq q^2\leq6\ \text{GeV}^2$) and a high $q^2$ bin ($14.18\leq q^2\leq16\ \text{GeV}^2$) for each of the following 4 CP-even $B\to K^*\mu\mu$ angular observables: $F_\text{L}, P_2, P_4'$, and $P_5'$.  We did not include other CP-even observables because the theoretical uncertainty of those observables are much too big.  Hence, they do not constrain our model.  These angular observables are calculated by \texttt{superiso} version 3.4 \cite{Mahmoudi:2008tp}.  Since \texttt{superiso} assumes that all soft parameters are real and only takes the diagonal entries of the trilinear couplings into account, we do not include CP-odd $B\to K^*\mu\mu$ angular observables in our analysis.

Table \ref{tab:obs} includes $45$ observables that we include in our global $\chi^2$ analysis.  The program that calculates these observables and the theoretical errors that are assigned to each observable are also included in the table.  The theoretical uncertainty for the $B\to K^*\mu\mu$ observables are taken from the \texttt{superiso} manual.  However, since \texttt{superiso} does not take into account the imaginary part of the soft parameters, we assigned an additional $15\%$ theoretical errors to the calculation.

To perform the global $\chi^2$ analysis, we construct a $\chi^2$ function
\begin{align}
  \chi^2 = \sum_{i} \frac{|x_i^\text{th} - x_i^\text{exp}|^2}{\sigma_i^2} \,,
\end{align}
where $x_i^\text{th}$ are the calculated values, $x_i^\text{exp}$ are the experimentally measured values, and $\sigma_i^2$ are the sum of the squares of the experimental and theoretical uncertainties, which are also listed in Table \ref{tab:obs}.  To find the minimum of this $\chi^2$ function, we use the \texttt{Minuit} package maintained by CERN \cite{James:1975dr}.  As in most minimization problems, obtaining the true global minimum is not guaranteed.  To increase the likelihood of obtaining the true global minimum, we iterate the minimization process with random initial guesses for the free parameters.

To calculate the $\chi^2$/d.o.f, we assume that the observables are uncorrelated.  So, for the universal boundary condition, we have $45-24=21$ degrees of freedom, while for the mirage boundary condition, we have $45-25=20$ degrees of freedom.  Given these gross assumptions, one should not take the value of $\chi^2$/d.o.f too seriously.

\begin{table}[!htbp]\footnotesize
\begin{center}
\scalebox{0.7}{
\renewcommand{\arraystretch}{1.2}
\begin{tabular}{|l|l|l|l|c|}
\hline
\textbf{Observable} &  \textbf{Exp.~Value}   & \textbf{Ref.}          & \textbf{Program} &  \textbf{Th.~Error}  \\
\hline
\hline
$M_Z$ & $91.1876\pm0.0021\ \text{GeV}$ & \cite{Agashe:2014kda} & Input & $0.0\%$ \\
$M_W$ & $80.385\pm0.015\ \text{GeV}$ & \cite{Agashe:2014kda} & \texttt{maton} & $0.5\%$ \\
$\alpha_\text{em}$ & $1/137.035999074(44)$ & \cite{Agashe:2014kda} & \texttt{maton} & $0.5\%$ \\
$G_\mu$ & $1.1663787(6)\times10^{-5}\ \text{GeV}^{-2}$ & \cite{Agashe:2014kda} & \texttt{maton} & $1\%$ \\
$\alpha_3(M_Z)$ & $0.1185\pm0.0006$ & \cite{Agashe:2014kda} & \texttt{maton} & $0.5\%$ \\
\hline
$M_t$ & $173.21\pm0.51\pm0.71\ \text{GeV}$ & \cite{Agashe:2014kda} & \texttt{maton} & $0.5\%$ \\
$m_b(m_b)$ & $4.18\pm0.03\ \text{GeV}$ & \cite{Agashe:2014kda} & \texttt{maton} & $0.5\%$ \\
$M_\tau$ & $1776.82\pm0.16\text{ MeV}$ & \cite{Agashe:2014kda} & \texttt{maton} & $0.5\%$ \\
\hline
$m_b-m_c$ & $3.45\pm0.05\ \text{GeV}$ & \cite{Agashe:2014kda} & \texttt{maton} & $10\%$ \\
$m_c(m_c)$ & $1.275\pm0.025\ \text{GeV}$ & \cite{Agashe:2014kda} & \texttt{maton} & $0.5\%$ \\
$m_s(2\ \text{GeV})$ & $95\pm5\text{ MeV}$ & \cite{Agashe:2014kda} & \texttt{maton} & $0.5\%$ \\
$m_s/m_d\,(2\ \text{GeV})$ & $17-22$ & \cite{Agashe:2014kda} & \texttt{maton} & $0.5\%$ \\
$Q$ & $21-25$ & \cite{Agashe:2014kda} & \texttt{maton} & $5\%$ \\
$M_\mu$ & $105.6583715(35)\text{ MeV}$ & \cite{Agashe:2014kda} & \texttt{maton} & $0.5\%$ \\
$M_e$ & $0.510998928(11)\text{ MeV}$ & \cite{Agashe:2014kda} & \texttt{maton} & $0.5\%$ \\
\hline
$|V_{us}|$ & $0.2253\pm0.0008$ & \cite{Agashe:2014kda} & \texttt{maton} & $0.5\%$ \\
$|V_{cb}|$ (Inclusive) & $0.0422\pm0.0007$ & \cite{Agashe:2014kda} & \texttt{maton} & $0.5\%$ \\
$|V_{cb}|$ (Exclusive) & $0.0395\pm0.0008$ & \cite{Agashe:2014kda} & \texttt{maton} & $0.5\%$ \\
$|V_{cb}|$ (Both) & $0.0408\pm0.0021$ & \cite{Agashe:2014kda} & \texttt{maton} & $0.5\%$ \\
$|V_{ub}|$ (Inclusive) & $0.00441\pm0.00024$ & \cite{Agashe:2014kda} & \texttt{maton} & $0.5\%$ \\
$|V_{ub}|$ (Exclusive) & $0.00328\pm0.00029$ & \cite{Agashe:2014kda} & \texttt{maton} & $0.5\%$ \\
$|V_{ub}|$ (Both) & $0.00385\pm0.00086$ & \cite{Agashe:2014kda} & \texttt{maton} & $0.5\%$ \\
$|V_{td}|$ & $0.00840\pm0.0006$ & \cite{Agashe:2014kda} & \texttt{maton} & $0.5\%$ \\
$|V_{ts}|$ & $0.0400\pm0.0027$ & \cite{Agashe:2014kda} & \texttt{maton} & $0.5\%$ \\
$\sin2\beta$ & $0.682\pm0.019$ & \cite{Agashe:2014kda} & \texttt{maton} & $0.5\%$ \\
$\epsilon_K$ & $(2.2325\pm0.0155)\times 10^{-3}$ & \cite{Agashe:2014kda} & \texttt{susyflavor}\cite{Crivellin:2012jv} & $10\%$ \\
\hline
$\Delta m_{B_s}/\Delta m_{B_d}$ & $35.0345\pm0.3884$ & \cite{Agashe:2014kda} & \texttt{susyflavor}\cite{Crivellin:2012jv} & $20\%$ \\
$\Delta m_{B_d}$ & $(3.337\pm0.033)\times 10^{-10}\text{ MeV}$ & \cite{Agashe:2014kda} & \texttt{susyflavor}\cite{Crivellin:2012jv} & $20\%$ \\
\hline
$\Delta m_{21}^2$ & $(7.02-8.09)\times10^{-5}\text{ eV}^2$ (3$\sigma$ range) & \cite{Gonzalez-Garcia:2014bfa} & \texttt{maton} & $0.5\%$ \\
$\Delta m_{31}^2$ & $(2.317-2.607)\times10^{-3}\text{ eV}^2$ (3$\sigma$ range) & \cite{Gonzalez-Garcia:2014bfa} & \texttt{maton} & $0.5\%$ \\
$\sin^2\theta_{12}$ & $0.270-0.344$ (3$\sigma$ range) & \cite{Gonzalez-Garcia:2014bfa} & \texttt{maton} & $0.5\%$ \\
$\sin^2\theta_{23}$ & $0.382-0.643$ (3$\sigma$ range) & \cite{Gonzalez-Garcia:2014bfa} & \texttt{maton} & $0.5\%$ \\
$\sin^2\theta_{13}$ & $0.0186-0.0250$ (3$\sigma$ range) & \cite{Gonzalez-Garcia:2014bfa} & \texttt{maton} & $0.5\%$ \\
\hline
$M_h$ & $125.7\pm0.4\ \text{GeV}$ & \cite{Agashe:2014kda} & \texttt{splitsuspect}\cite{Bernal:2007uv} & $3$ GeV \\
\hline
$\text{BR}(b\rightarrow s\gamma)$ & $(343\pm21\pm7) \times 10^{-6}$ & \cite{Amhis:2014hma} & \texttt{superiso}\cite{Mahmoudi:2008tp} & $40\%$ \\
$\text{BR}(B_s\rightarrow\mu^+\mu^-)$ & $(2.8^{+0.7}_{-0.6})\times10^{-9}$ & \cite{CMS:2014xfa} & \texttt{susyflavor}\cite{Crivellin:2012jv} & $20\%$ \\
$\text{BR}(B_d\rightarrow\mu^+\mu^-)$ & $(3.9^{+1.6}_{-1.4})\times10^{-10}$ & \cite{CMS:2014xfa} & \texttt{susyflavor}\cite{Crivellin:2012jv} & $20\%$ \\
$\text{BR}(B\rightarrow\tau\nu)$ & $(114\pm22)\times10^{-6}$ & \cite{Amhis:2014hma} & \texttt{susyflavor}\cite{Crivellin:2012jv} & $50\%$ \\
\hline
$\text{BR}(B \rightarrow K^*\mu^+\mu^-)_{1\leq q^2\leq 6\ \text{GeV}^2}$  &  $0.34\pm0.03\pm0.04\pm0.02^{+0.00}_{-0.03}\times10^{-7}$ & \cite{Aaij:2013iag} & \texttt{superiso}\cite{Mahmoudi:2008tp} & $105\%$ \\
$\text{BR}(B \rightarrow K^*\mu^+\mu^-)_{14.18\leq q^2\leq 16\ \text{GeV}^2}$ & $0.45\pm0.06\pm0.04\pm0.04^{+0.00}_{-0.05}\times10^{-7}$ & \cite{Aaij:2013iag} & \texttt{superiso}\cite{Mahmoudi:2008tp} & $190\%$ \\
$q_0^2(\text{A}_\text{FB}(B \rightarrow K^*\mu^+\mu^-))$     &  $4.9\pm0.9 \text{GeV}^2$ & \cite{Aaij:2013iag} & \texttt{superiso}\cite{Mahmoudi:2008tp} & $25\%$ \\
$F_L(B \rightarrow K^*\mu^+\mu^-)_{1\leq q^2\leq 6\ \text{GeV}^2}$ & $0.65^{+0.08+0.03}_{-0.07-0.03}$ & \cite{Aaij:2013iag} & \texttt{superiso}\cite{Mahmoudi:2008tp} & $45\%$ \\
$F_L(B \rightarrow K^*\mu^+\mu^-)_{14.18\leq q^2\leq 16\ \text{GeV}^2}$ & $0.33^{+0.08+0.03}_{-0.07-0.03}$ & \cite{Aaij:2013iag} & \texttt{superiso}\cite{Mahmoudi:2008tp} & $80\%$ \\
$-2P_2=A_T^\text{Re}(B \rightarrow K^*\mu^+\mu^-)_{1\leq q^2\leq 6\ \text{GeV}^2}$ & $-0.66^{+0.24+0.04}_{-0.22-0.01}$ & \cite{Aaij:2013iag} & \texttt{superiso}\cite{Mahmoudi:2008tp} & $95\%$ \\
$-2P_2=A_T^\text{Re}(B \rightarrow K^*\mu^+\mu^-)_{14.18\leq q^2\leq 16\ \text{GeV}^2}$ & $1.00^{+0.00+0.00}_{-0.05-0.02}$ & \cite{Aaij:2013iag} & \texttt{superiso}\cite{Mahmoudi:2008tp} & $45\%$ \\
$P_4'(B \rightarrow K^*\mu^+\mu^-)_{1\leq q^2\leq 6\ \text{GeV}^2}$ & $0.58^{+0.36}_{-0.32}\pm0.06$ & \cite{Aaij:2013qta} & \texttt{superiso}\cite{Mahmoudi:2008tp} & $30\%$ \\ 
$P_4'(B \rightarrow K^*\mu^+\mu^-)_{14.18\leq q^2\leq 16\ \text{GeV}^2}$ & $-0.18^{+0.70}_{-0.54}\pm0.08$ & \cite{Aaij:2013qta} & \texttt{superiso}\cite{Mahmoudi:2008tp} & $35\%$ \\ 
$P_5'(B \rightarrow K^*\mu^+\mu^-)_{1\leq q^2\leq 6\ \text{GeV}^2}$ & $0.21^{+0.20}_{-0.21}\pm0.03$ & \cite{Aaij:2013qta} & \texttt{superiso}\cite{Mahmoudi:2008tp} & $45\%$ \\
$P_5'(B \rightarrow K^*\mu^+\mu^-)_{14.18\leq q^2\leq 16\ \text{GeV}^2}$ & $-0.79^{+0.20}_{-0.13}\pm0.18$ & \cite{Aaij:2013qta} & \texttt{superiso}\cite{Mahmoudi:2008tp} & $60\%$ \\
\hline
\end{tabular}
}
\end{center}
\caption{\small This table includes 45 observables that we fit.  All experimental errors are $1\sigma$ unless otherwise indicated.  Column 4 shows the software package that gives us the theoretical prediction.  $M_Z$ is fit precisely to impose electroweak symmetry breaking.  To account for the inconsistencies in the inclusive and exclusive measurements of $|V_{ub}|$ and $|V_{cb}|$, we perform the global $\chi^2$ analysis using the inclusive and exclusive measurements separately.  We also perform an additional analysis where the error bars of $|V_{ub}|$ and $|V_{cb}|$ cover both the inclusive and exclusive values.  In additional to the theoretical errors indicated in \texttt{superiso} manual, we added an additional 15\% error to each of the $B\to K^*\mu^+\mu^-$ observables because \texttt{superiso} does not take into account the phases of soft terms.}
\label{tab:obs}
\end{table}

\section{Results: Global $\chi^2$ Analysis}
\label{sec:global_chi2}

Several benchmark points with the results of the global $\chi^2$ analysis are given in Appendix \ref{app:benchmark} and a
plot of $\chi^2$ as a function of the parameter $m_{16}$ is given in Figure \ref{fig:chi2_line_plot} or $\chi^2$ contours in the two dimensional
plane of $M_{\tilde g}$ vs. $m_{16}$ is given in Figure \ref{fig:contour}.  Let us now discuss some features of the results.

\subsection{Inclusive vs.~Exclusive $|V_{ub}|$ and $|V_{cb}|$}

Due to the discrepancy between the values of $|V_{ub}|$ and $|V_{cb}|$ determined from inclusive and exclusive semi-leptonic decay.  We define three different $\chi^2$ functions:
\begin{enumerate}
  \item $|V_{ub}|$ and $|V_{cb}|$ are taken to be the inclusive values
  \item $|V_{ub}|$ and $|V_{cb}|$ are taken to be the exclusive values
  \item $|V_{ub}|$ and $|V_{cb}|$ are taken to be the average of inclusive and exclusive values with error bars overlapping with the error bars of both the inclusive and the exclusive measurements
\end{enumerate}

\begin{figure}[h!]
  \begin{center}
    \includegraphics[width=0.8\textwidth]{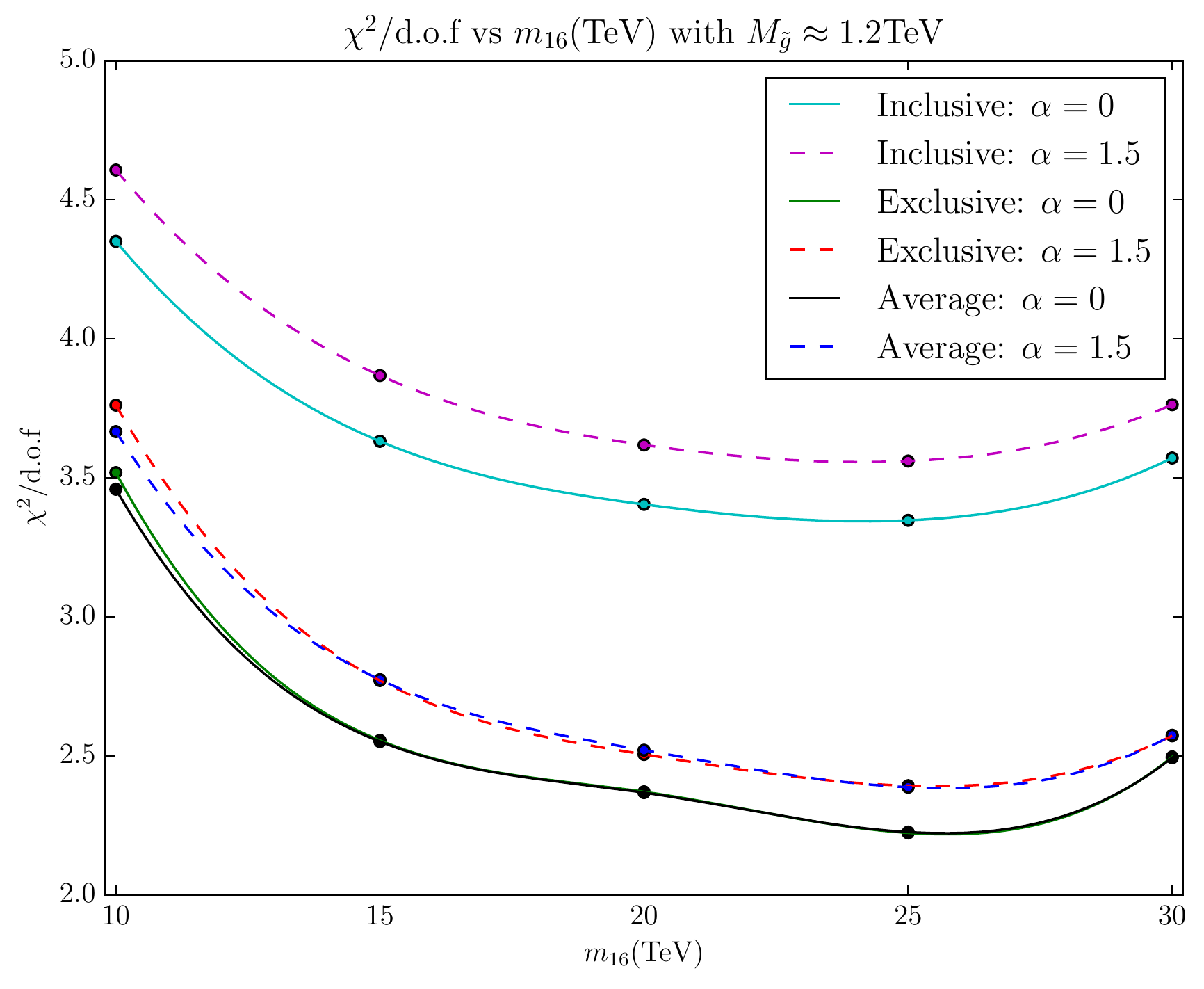}
    \caption{This plot shows the value of $\chi^2$/d.o.f as a function of $m_{16}$ for cases where the value of $|V_{ub}|$ and $|V_{cb}|$ are taken to be the inclusive values, the exclusive values, or the average of inclusive and exclusive values.  Solid lines refers to the universal boundary condition, $\alpha=0$, while dashed lines refer to the mirage boundary condition with $\alpha=1.5$.  This plot shows that our model favors the exclusive values of $|V_{ub}|$ and $|V_{cb}|$.}
    \label{fig:chi2_line_plot}
  \end{center}
\end{figure}

The results of these three analyses are shown in Figure \ref{fig:chi2_line_plot}.  We see that for both the universal boundary condition $\alpha=0$ and mirage boundary condition with $\alpha=1.5$, the $\chi^2$/d.o.f obtained by fitting to the inclusive values are the biggest.  Hence, we predict that the exclusive values of $|V_{ub}|$ and $|V_{cb}|$ are the correct values for both universal and mirage gaugino masses.

Since the $\chi^2$ difference between case (2) and case (3) is small and to be conservative, the analyses of the rest of the paper are done for case (3), where $|V_{ub}|$ and $|V_{cb}|$ are the average of the inclusive and exclusive values.

\subsection{SUSY Non-decoupled observables}
\label{ssec:b_physics}

\vspace{.1in}
{\em B physics observables}
\vspace{.1in}

Some of the measured angular observables of $B\to K^*\mu^+\mu^-$ are in tension with the SM prediction.  For example, $P_4'$ in the high $q^2$ bin ($14.18\leq q^2\leq16\ \text{GeV}^2$) has a $2.7\sigma$ discrepancy with the SM prediction, $P_5'$ in the low $q^2$ bin ($1\leq q^2\leq 6\ \text{GeV}^2$) has a $2.5\sigma$ discrepancy with the SM prediction, and $P_2$ in the low $q^2$ bin has a $2\sigma$ discrepancy with the SM prediction \cite{Aaij:2013qta, Aaij:2013iag, Altmannshofer:2013foa}.  These observables are defined in \cite{DescotesGenon:2012zf, Matias:2012xw}. In addition, previous analysis \cite{Altmannshofer:2013foa, Descotes-Genon:2013wba} found that the tension in $P_4'$ of the high $q^2$ bin cannot be explained by the MSSM.  On the other hand, the tension of $F_L$ and $P_5'$ of the low $q^2$ bin can be explained by the MSSM by having a negative contribution to the $C_7$ Wilson coefficient.  In the standard model $C_7\approx-0.32$.  The tension in $F_L$ and $P_5'$ can be further reduced by making $C_7$ more negative \cite{Altmannshofer:2013foa, Mahmoudi:2014mja}.

In the MSSM, chargino-stop loops and charged Higgs loop contribute to $C_7$.  The $C_7$ contribution from the charged Higgs is always negative.  The charged Higgs of our model has mass around $2\ \text{TeV}$.  So, the charged Higgs contribution to $C_7$ is non-negligible and is in the correct direction.

The chargino-stop loop contribution of $C_7^\text{MSSM}$ has the following form \cite{Albrecht:2007ii}\footnote{Note, the equation for the chargino contribution to $C_7^\text{MSSM}$ given in Eqn. 21, Ref. \cite{Altmannshofer:2013foa} apparently has the wrong sign.}
\begin{align}
  C_7^\text{MSSM} = \frac{\mu A_t\tan\beta}{m_{\tilde t}^4}\text{sign}(C_7^\text{SM}) \,.
\end{align}
Since $\text{sign}(\mu A_t)$ is negative in our model, this term contributes to $C_7$ in the wrong direction.  Hence, to reduce the contribution of this term, our model favors large scalar masses.

From our global $\chi^2$ analysis, we see that the calculated value of $P_4'$ in the high $q^2$ bin does not depend on $m_{16}$, which is expected.  In addition, the value of $P_4'$ calculated in our model is in agreement with the SM.  Hence, our results are in agreement with previous analysis that the tension in $P_4'$ cannot be explained in the MSSM.  As shown in Table \ref{tab:trend}, the tension of $F_L$ and $P_5'$ with the experimental values decreases as $m_{16}$ increases.  This is again in agreement with our expectation as explained above.

\vspace{.1in}
{\em SUSY corrections to the W mass}
\vspace{.1in}

On the other hand, the correction for $M_W$ is given by \cite{Djouadi:1998sq,Heinemeyer:2004gx,Barger:2012hr}
\begin{align}
  \delta M_W \approx \frac{M_W}{2}\frac{c_W^2}{c_W^2-s_W^2}\Delta\rho
\end{align}
and the 1-loop squark contribution is given by
\begin{align}
  \Delta\rho_1^\text{SUSY} &= \frac{3G_\mu}{8\sqrt{2}\pi^2}[-s_{\tilde t}^2c_{\tilde t}^2F_0(m_{\tilde t_1}^2,m_{\tilde t_2}^2) - s_{\tilde b}^2c_{\tilde b}^2F_0(m_{\tilde b_1}^2,m_{\tilde b_2}^2)
  \\&\phantom{==} + c_{\tilde t}^2c_{\tilde b}^2F_0(m_{\tilde t_1}^2,m_{\tilde b_1}^2) + c_{\tilde t}^2s_{\tilde b}^2F_0(m_{\tilde t_1}^2,m_{\tilde b_2}^2) + s_{\tilde t}^2c_{\tilde b^2}F_0(m_{\tilde t_2}^2, m_{\tilde b_1}^2) + s_{\tilde t}^2s_{\tilde b}^2F_0(m_{\tilde t_2}^2,m_{\tilde b_2^2})]
\end{align}
where $s_W=\sin\theta_W, c_W=\cos\theta_W, s_{\tilde q}=\sin\theta_{\tilde q}, c_{\tilde q}=\cos\theta_{\tilde q}$, and
\begin{align}
  F_0(x,y) = x + y - \frac{2xy}{x-y}\ln\frac{x}{y} \,.
\end{align}
$F_0$ has properties of $F_0(x,x)=0$ and $F_0(x,0)=x$.  Hence, we see that when the mass splitting of the squarks is large, the SUSY contribution to the 1-loop $M_W$ can be significant.  This is in agreement with our analysis which shows that the pull from $M_W$ increases as the value of $m_{16}$ increases above 20 TeV.  Hence, SUSY corrections to $M_W$ are significant and they can go in the right direction.

\vspace{.1in}
{\em Light Higgs mass}
\vspace{.1in}

Fitting to the Higgs mass also constrains the value of $m_{16}$.  The dominant one-loop contribution to the Higgs mass is given by
\begin{align}
  m_h^2 \approx m_Z^2\cos^22\beta + \frac{3}{(4\pi)^2}\frac{m_t^4}{v^2}\left[\ln\frac{M_{SUSY}^2}{m_t^2}+\frac{X_t}{M_{SUSY}^2}\left(1-\frac{X_t^2}{12 M_{SUSY}^2}\right)\right]
\end{align}
where $X_t=A_t-\mu/\tan\beta$ is the stop mixing parameter and $M_{SUSY}^2 = m_{\tilde t_1} \  m_{\tilde t_2}$. In our model, $X_t < -\sqrt{6} \ M_{SUSY}$ and the ratio $X_t/M_{SUSY}$ becomes less negative as $m_{16}$ increases. Hence, as $m_{16}$ increases, the Higgs mass also increases.  The pull in $\chi^2$ due to $M_h$ has a minimum around $m_{16} = 25$ TeV.

\begin{figure}[h!]
  \begin{center}
    \includegraphics[width=0.8\textwidth]{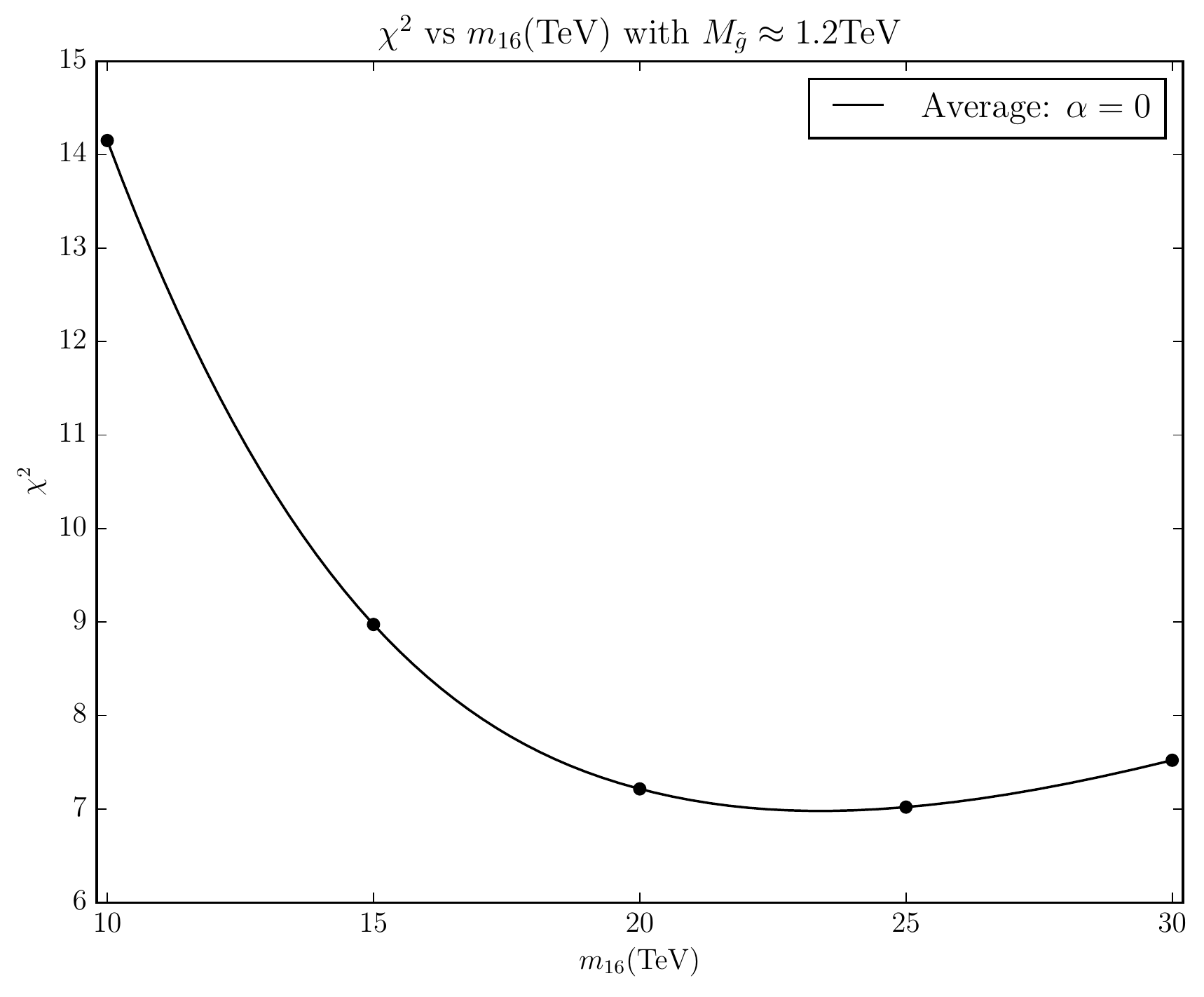}
    \caption{This figure shows the contribution to $\chi^2$ as a function of $m_{16}$ just from the set of observables listed in Table \ref{tab:trend}.  Fitting to the values of $M_W$, $M_h$, and $b$-physics observables listed in Table \ref{tab:trend} helps explain why $\chi^2$ is minimized at $m_{16} \approx 25\ \text{TeV}$.}
    \label{fig:trend}
  \end{center}
\end{figure}

Hence,  the contributions of $M_W$, $M_h$, and $b$-physics observables to $\chi^2$, as listed in Table \ref{tab:trend}, help explain the shape of $\chi^2$ as a function of $m_{16}$ (see Figure \ref{fig:trend}).

\begin{table}[!htbp]
  \begin{center}
    \begin{tabular}{| l | r | r | r | r | r |}
      \hline & \multicolumn{5}{c|}{Pull} \\
      \cline{2-6} $m_{16}$ & 10 & 15 & 20 & 25 & 30 \\
      \hline $M_W$ & 0.2110 & 0.1878 & 0.1851 & 0.2320 & 0.3981 \\
      \hline $M_h$ & 2.5474 & 1.1795 & 0.3454 & 0.1882 & 0.6582 \\
      \hline $BR(B\to\tau\nu)$ & 1.1978 & 1.3952 & 1.3557 & 1.3588 & 1.3771 \\
      \hline $F_L(B\to K^*\mu^+\mu^-)_{1\leq q^2\leq6\ \text{GeV}^2}$ & 0.2696 & 0.2488 & 0.2219 & 0.2101 & 0.2057 \\
      $P_4'(B\to K^*\mu^+\mu^-)_{1\leq q^2\leq6\ \text{GeV}^2}$ & 1.7066 & 1.7066 & 1.7066 & 1.7066 & 1.7066 \\
      $P_5'(B\to K^*\mu^+\mu^-)_{1\leq q^2\leq6\ \text{GeV}^2}$ & 2.4110 & 2.3432 & 2.2746 & 2.2451 & 2.2339 \\
      \hline $\chi^2$ & 14.1511 & 8.9744 & 7.2154 & 7.0206 & 7.5220 \\
      \hline
    \end{tabular}

    \vspace{0.5cm}
    {\footnotesize
    \begin{tabular}{| l | r | r | r | r | r | r |}
      \hline & \multicolumn{5}{c|}{Fit Value} & Exp. \\
      \cline{2-6} $m_{16}$ & 10 & 15 & 20 & 25 & 30 & Value \\
      \hline $M_W$ & 80.4699 & 80.4606 & 80.4595 & 80.4784 & 80.5454 & 80.3850 \\
      \hline $M_h$ & 117.9901 & 122.1303 & 124.6547 & 126.2697 & 127.6920 & 125.7000 \\
      \hline $BR(B\to\tau\nu)\times10^5$ & 6.6329 & 6.1340 & 6.2299 & 6.2223 & 6.1778 & 11.4000 \\
      \hline $F_L(B\to K^*\mu^+\mu^-)_{1\leq q^2\leq6\ \text{GeV}^2}$ & 0.7434 & 0.7353 & 0.7251 & 0.7207 & 0.7191 & 0.6500 \\
      $P_4'(B\to K^*\mu^+\mu^-)_{1\leq q^2\leq6\ \text{GeV}^2}$ & 0.8174 & 0.6711 & 0.5921 & 0.5717 & 0.5657 & 0.5800 \\
      $P_4'(B\to K^*\mu^+\mu^-)_{14.18\leq q^2\leq16\ \text{GeV}^2}$ & 1.2190 & 1.2190 & 1.2190 & 1.2190 & 1.2190 & -0.1800 \\
      $P_5'(B\to K^*\mu^+\mu^-)_{1\leq q^2\leq6\ \text{GeV}^2}$ & -0.7301 & -0.5529 & -0.4625 & -0.4335 & -0.4235 & 0.2100 \\
      \hline
    \end{tabular}
    }
    \caption{This table shows the set of observables with $\alpha=0$ and $M_{\tilde g}\approx1.2\ \text{TeV}$.  As argued by Altmannshofer et.~al.  $B\to K^*\mu^+\mu^-$ favors large $m_{\tilde t}$ \cite{Altmannshofer:2013foa}, which is in agreement with our analysis.  On the other hand, as explained in the text, fitting to $M_W$ and $M_h$ disfavors large $m_{\tilde t}$.  These effects collectively contribute to having a minimum $\chi^2$ around $m_{16}=25\ \text{TeV}$. The $\chi^2$ value is plotted as a function of $m_{16}$ in Figure \ref{fig:trend}.  In addition to the observables contributing directly to a minimum of $\chi^2$ at $m_{16}=25\ \text{TeV}$, also included in this table is the calculated value of $P_4'$ at high $q^2$ bin.  This is to illustrate that the value of $P_4'$ does not depend on the $m_{16}$, which again agrees with previous analysis~\cite{Altmannshofer:2013foa, Descotes-Genon:2013wba}.}
    \label{tab:trend}
  \end{center}
\end{table} 

\newpage
\subsection{Bounds on $M_{\tilde g}$}
\label{ssec:gluino}

To obtain a better picture for the favored value of gluino mass, we plotted two contour plots of $M_{\tilde g}$ vs. $m_{16}$.  One for the universal boundary condition $\alpha=0$ and another for the mirage boundary condition with $\alpha=1.5$.  The current bound on $M_{\tilde g}$, for our model, is around $1.2\ \text{TeV}$ \cite{Anandakrishnan:2014nea}.  The contour plots are created by calculating $\chi^2$/d.o.f for 25 equally distributed values of $m_{16}$ and $M_{1/2}$, which gives us $1 < M_{\tilde g}  <2\ \text{TeV}$.  We then use cubic interpolation to obtain the smooth contours of $\chi^2$/d.o.f.

In addition to the contour lines of the $\chi^2$/d.o.f, we also plotted a $4\sigma$ contour line.  From this, we see that for mirage boundary conditions $M_{\tilde g} \lesssim 1.8\ \text{TeV}$.  However, for universal boundary condition, the $4\sigma$ $M_{\tilde g}$ bound can be as high as $3\ \text{TeV}$, which is not shown in the Figure.  Hence, for mirage boundary conditions, we expect the $4\sigma$ bound on the gluino mass to be within reach in the next run of the LHC.  As pointed out by \cite{Anandakrishnan:2014nea}, the dominant decay mode of the gluino in the universal gaugino mass boundary condition is $t \ b \ \tilde\chi_1^\mp$.  The remaining decay modes are $t \ \bar t \ \tilde\chi_i^0$ and $b \ \bar b \ \chi_i^0$ for $i=1,2$.  On the other hand, the dominant mode for gluino decay in the mirage gaugino mass boundary condition is $t \ b \ \tilde\chi_i^\mp$ for $i=1,2$.  In all cases, the dominant signature for gluinos in this model is given by $b$ jets, leptons and missing $E_T$ \cite{Anandakrishnan:2014nea}.

\begin{figure}[!htbp]
  \begin{center}
    \includegraphics[width=0.45\textheight]{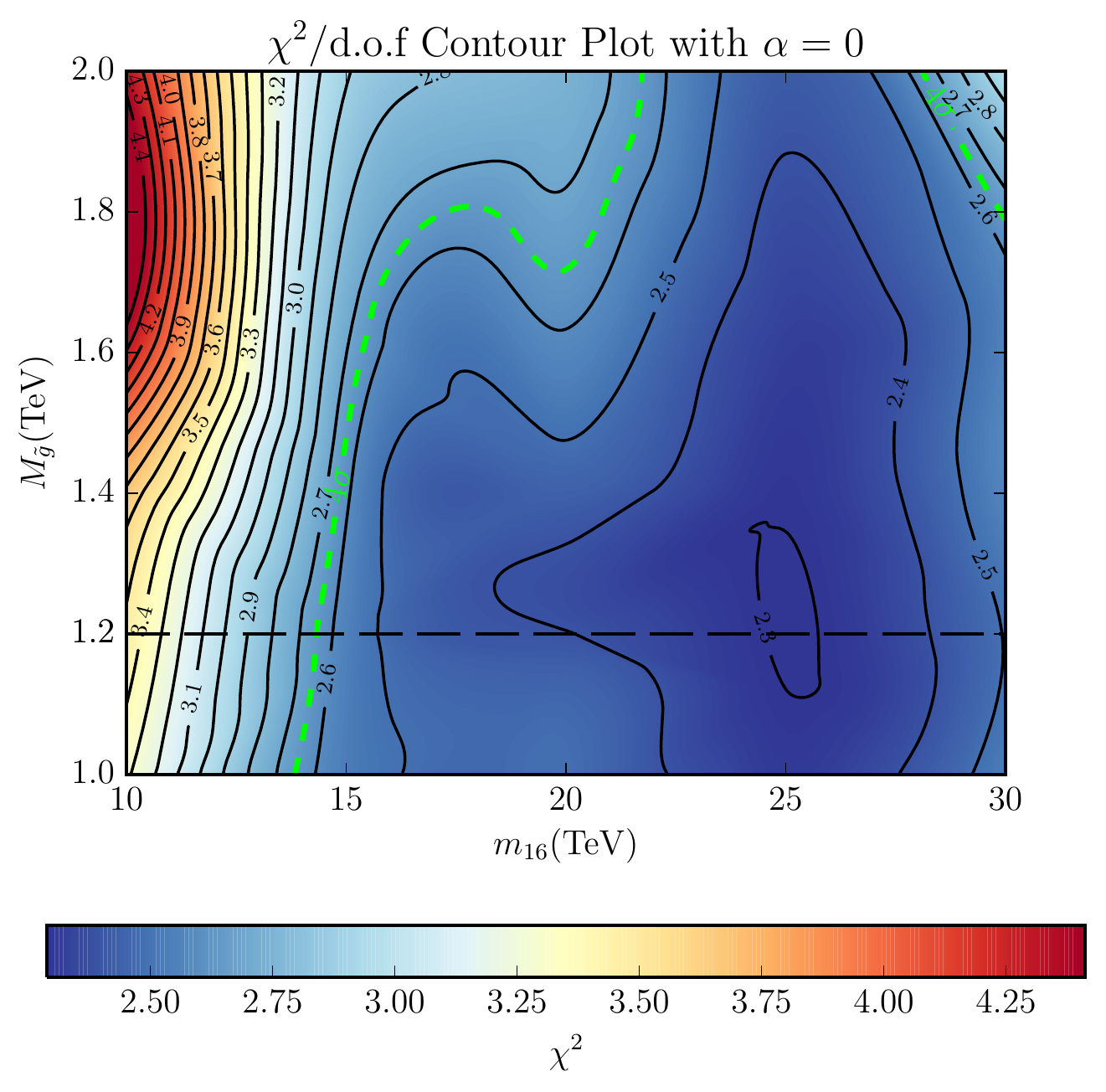}
    \includegraphics[width=0.45\textheight]{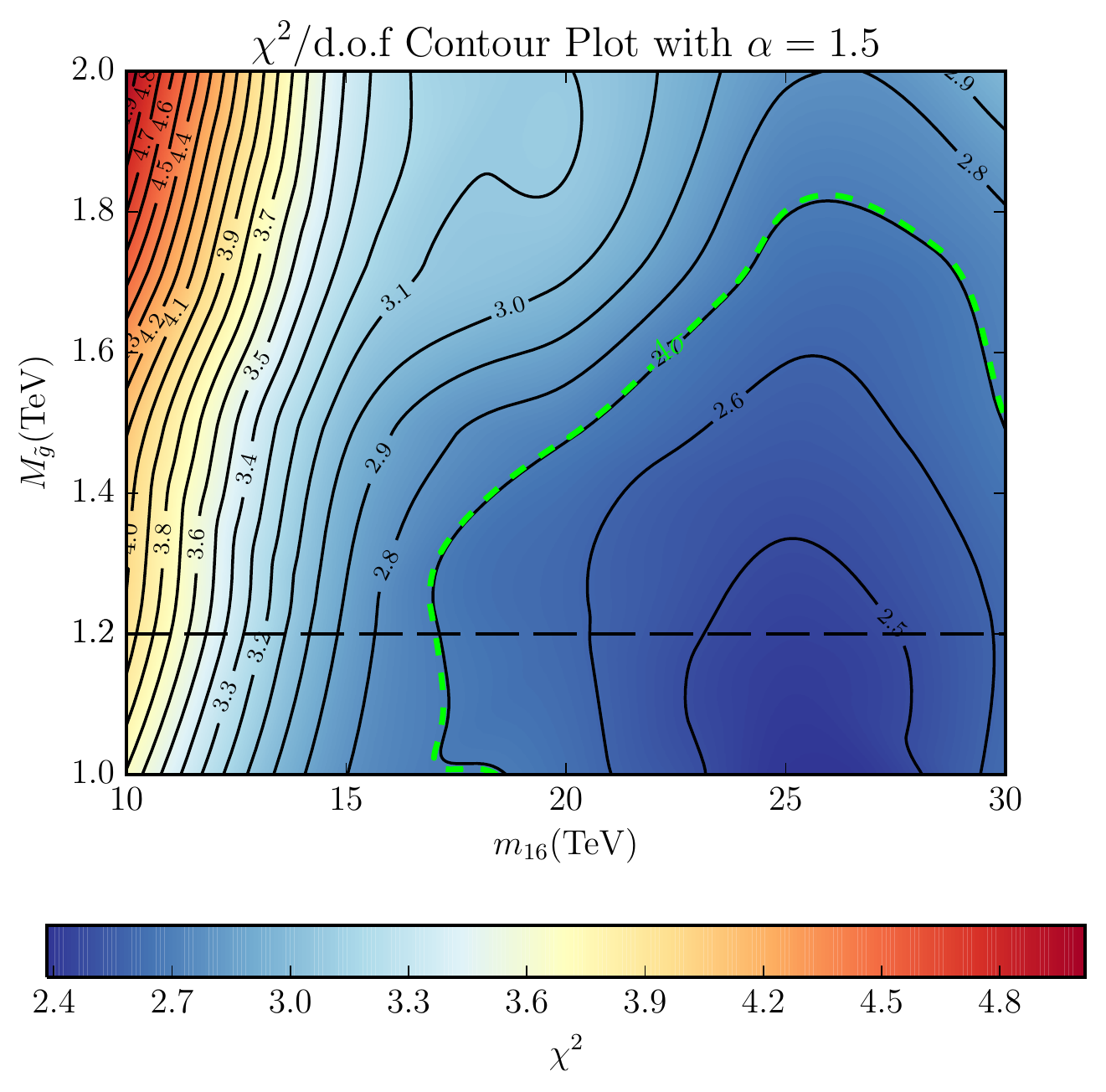}
    \caption{These plots show the contour of $\chi^2$/d.o.f as a function of the $M_{\tilde g}$ and $m_{16}$.  The $4\sigma$ bound is also included in the plot.  For $\alpha=1.5$, the upper bound is within reach of the next run of LHC.  In addition, we also see that our model favors $m_{16} \approx 25\ \text{TeV}$.}
    \label{fig:contour}
  \end{center}
\end{figure}

\subsection{Additional predictions for some benchmark points.}
\label{ssec:predictions}

The mass spectrum of the benchmark point of $M_{\tilde g} \approx 1.2\ \text{TeV}$ and $m_{16} = 25\ \text{TeV}$ is shown in Table \ref{tab:prediction}.  From the $\chi^2$ analysis, we see that the scalar masses are predicted to be around $5\ \text{TeV}$, while the first and second generation scalars have mass around $m_{16} \approx 25\ \text{TeV}$.  With scalars in this mass range, the stop in our model does not completely decouple and can have non-negligible effects on flavor physics.  In addition, our light Higgs is SM-like with the heavy Higgs with mass around $2\ \text{TeV}$.

In Table \ref{tab:prediction}, we give the light sparticle masses, the CP violating angle for neutrino oscillations, $\delta$, the branching ratio $BR(\mu \to e \gamma)$ and the electric dipole moment of the electron for two different values of $M_{\tilde g}$ and for $\alpha = 0 \text{and} 1.5$.  Note that, in general, the gauginos are the lightest sparticles.  In addition, $BR(\mu \to e \gamma)$ and the electric dipole moment of the electron are within reach of future experiments.

\begin{table}[!htbp]
  \begin{center}
    \begin{tabular}{| l | r | r | r | r |}
      \hline $m_{16}$ & 25 & 25 & 25 & 25 \\
      $\alpha$ & 0 & 1.5 & 0 & 1.5 \\
      \hline $\chi^2$/d.o.f & 2.158 & 2.275 & 2.220 & 2.505 \\
      \hline $m_{\tilde t_1}$ & 4.903 & 5.011 & 4.909 & 5.249 \\
      $m_{\tilde t_2}$ & 6.021 & 6.120 & 6.033 & 6.301 \\
      $m_{\tilde b_1}$ & 5.989 & 6.088 & 6.455 & 6.606 \\
      $m_{\tilde b_2}$ & 6.454 & 6.541 & 6.445 & 6.267 \\
      $m_{\tilde\tau_1}$ & 9.880 & 9.931 & 9.912 & 10.040 \\
      $m_{\tilde\tau_2}$ & 15.369 & 15.365 & 15.393 & 15.516 \\
      $M_{\tilde g}$ & 1.202 & 1.187 & 1.613 & 1.690 \\
      $m_{\tilde\chi_1^0}$ & 0.203 & 0.551 & 0.279 & 0.900 \\
      $m_{\tilde\chi_2^0}$ & 0.404 & 0.665 & 0.538 & 1.018 \\
      $m_{\tilde\chi_1^+}$ & 0.404 & 0.665 & 0.538 & 1.018 \\
      $m_{\tilde\chi_2^+}$ & 1.128 & 1.243 & 1.232 & 1.537 \\
      $M_A$ & 2.194 & 2.082 & 2.477 & 3.352 \\
      \hline $\sin\delta$ & -0.289 & -0.482 & -0.520 & -0.576 \\
      $BR(\mu\to e\gamma)\times10^{13}$ & 1.108 & 1.430 & 1.239 & 1.340 \\
      $\text{edm}_e\times10^{30}(\text{e cm})$ & -1.403 & -3.305 & -1.763 & -5.886 \\
      \hline
    \end{tabular}
    \caption{Predictions with $m_{16}=25\ \text{TeV}$ for $M_{\tilde g} \approx 1.2\ \text{TeV}$ and $1.6\ \text{TeV}$.  All masses in the table are in TeV units.  Our prediction for the branching ratio $\mu\to e\gamma$ is consistent with the current upper bound of $5.7 \times 10^{-13}$ \cite{Agashe:2014kda}.   In addition, our prediction of the electron electric dipole moment is consistent with the current upper bound of $10.5 \times10^{-28}\ \text{e cm}$ \cite{Agashe:2014kda}.}
    \label{tab:prediction}
  \end{center}
\end{table}

\section{Results: Fine-Tuning}
\label{sec:fine_tuning}

We studied the fine-tuning of our model using the fine-tuning measure introduced by Ellis et. al.~\cite{Ellis:1986yg}, and studied in detail by Barbieri and Giudice \cite{Barbieri:1987fn},
\begin{align}
  \Delta_\text{BG} = \max\Delta_{a_i} \,,
  \phantom{==}
  \Delta_{a_i} = \left|\frac{\partial\ln M_Z^2}{\partial\ln a_i^2}\right| \,,
  \label{eq:ft_bg}
\end{align}
where $a_i$s are input parameters of the model.  This fine-tuning measure calculates the sensitivity of $M_Z$ due to a small variation of the input parameters defined at GUT scale.

Electroweak symmetry is broken radiatively in our model.  From radiative electroweak symmetry breaking, the CP-odd Higgs mass, $m_A$, and the $\mu$-term are calculated at one-loop \cite{Pierce:1996zz}.  This calculation requires the physical $Z$ pole mass, $M_Z$.  Hence, in our model, $M_Z$ is fit precisely.  To make sure that radiative electroweak symmetry breaking is consistent, $m_A$ and $\mu$ are calculated iteratively until they converge.

On the other hand, when we calculate fine-tuning using (\ref{eq:ft_bg}), we use the benchmark points.  The benchmark points are the inputs that produce minimum $\chi^2$ value for their respective value of $m_{16}$ and $M_{1/2}$.  Hence, at each benchmark point, radiative electroweak symmetry breaking is consistent.  Thus, instead of fixing $M_Z$ and calculating $m_A$ and $\mu$ iteratively, we then use the value of $m_A$ and $\mu$ to calculate $M_Z$.  We then compare this value of $M_{Z}$ to the exact value to obtain the fine tuning parameter $\Delta_{BG}$.

The input parameters that we vary are $a_i=\{\mu, M_{1/2}, m_{16}, m_{H_u}, m_{H_d}, A_0\}$.  The results of our calculations are summarized in Table \ref{tab:ft}.  These results can be understood by the running of GUT scale parameters that contribute to $Z$ mass.  For $\tan\beta=10$, $M_Z$ written in terms of GUT scale parameters is \cite{Baer:2014ica,Feng:2013pwa,Martin:2007gf,Kobayashi:2006fh}
\begin{align}
  M_Z^2 \approx -2.18\mu^2 + 4.22M_{1/2}^2 - 0.82M_{1/2}A_0 + 0.22A_0^2 - 1.27m_{H_u}^2 - 0.053m_{H_d}^2 + 1.34m_{16}^2 \,.
\end{align}
Although the above equation is derived for $\tan\beta=10$, to the lowest order approximation, we do not expect this result to change drastically when $\tan\beta$ increases to $\approx 50$.  The calculated fine tuning values shown in Table \ref{tab:ft} are of the same order as the fine tuning predicted from this equation.  As an example, by direct substitution of the $m_{16}=20\ \text{TeV}$ benchmark points into the above equation, we find that if $m_{H_{u,d}}/m_{16}$ and $A_0/m_{16}$ are fixed, then $\Delta_{BG} \approx 2000$ is of the same order as our calculation.

\begin{table}[!htbp]
  \begin{center}
    \begin{tabular}{| l | r | r | r | r | r |}
      \multicolumn{6}{c}{Fine-Tuning of Benchmark Points with $\alpha=0$ and $M_{\tilde{g}}\approx1.2\ \text{TeV}$}
      \\ \hline
        & \multicolumn{5}{c|}{$m_{16}$}
      \\ \cline{2-6} Varying Parameters & 10TeV & 15TeV & 20TeV & 25TeV & 30TeV
      \\ \hline $\mu$ & 140 & 190 & 210 & 360 & 490
      \\ \hline $M_{1/2}$ & 260 & 340 & 400 & 430 & 450
      \\ \hline $m_{16}$ & 12000 & 27000 & 47000 & 74000 & 110000
      \\ \hline $m_{H_d}$ & 760 & 1500 & 3900 & 6100 & 8700
      \\ \hline $m_{H_u}$ & 10000 & 23000 & 40000 & 62000 & 89000
      \\ \hline $A_0$ & 9300 & 21000 & 39000 & 61000 & 85000
      \\ \hline\hline $m_{16}$ with $A_0/m_{16}$ fixed & 22000 & 49000 & 87000 & 130000 & 190000
      \\ \hline $m_{16}$ with $m_{H_u,d}/m_{16}$ fixed & 9500 & 22000 & 40000 & 62000 & 86000
      \\ \hline\hline $m_{16}$ with $m_{H_{u,d}}/m_{16}, A_0/m_{16}$ fixed & 240 & 400 & 630 & 740 & 850
      \\ \hline \multicolumn{6}{c}{}
      \\ \multicolumn{6}{c}{Fine-Tuning of Benchmark Points with $\alpha=1.5$ and $M_{\tilde{g}}\approx1.2\ \text{TeV}$}
      \\ \hline
        & \multicolumn{5}{c|}{$m_{16}$}
      \\ \cline{2-6} Varying Parameters & 10TeV & 15TeV & 20TeV & 25TeV & 30TeV
      \\ \hline $\mu$ & 110 & 240 & 290 & 340 & 380
      \\ \hline $M_{1/2}$ & 320 & 420 & 500 & 560 & 580
      \\ \hline $m_{16}$ & 12000 & 28000 & 48000 & 74000 & 110000
      \\ \hline $m_{H_d}$ & 750 & 1500 & 4600 & 6000 & 8500
      \\ \hline $m_{H_u}$ & 10000 & 23000 & 39000 & 62000 & 89000
      \\ \hline $A_0$ & 9200 & 21000 & 39000 & 60000 & 86000
      \\ \hline\hline $m_{16}$ with $A_0/m_{16}$ fixed & 22000 & 49000 & 87000 & 130000 & 190000
      \\ \hline $m_{16}$ with $m_{H_u,d}/m_{16}$ fixed & 9600 & 21000 & 39000 & 61000 & 87000
      \\ \hline\hline $m_{16}$ with $m_{H_{u,d}}/m_{16}, A_0/m_{16}$ fixed & 330 & 450 & 670 & 890 & 1100
      \\ \hline
    \end{tabular}
    \caption{Without fixing any ratios, the fine-tuning is 1 part in $10^5$. When the ratio of $m_{H_{u,d}}/m_{16}$ and $A_0/m_{16}$ are fixed, the fine-tuning according to (\ref{eq:ft_bg}) is about 1 part in $500$.  This suggests that these ratios should be fixed in a more fundamental natural theory.  In addition, fine-tuning increases as $m_{16}$ increases.  Hence, in our model, small $m_{16}$ is favored by naturalness.}
    \label{tab:ft}
  \end{center}
\end{table}

From Table \ref{tab:ft}, we see that if there are no constraints on the input parameters (first five rows), then the fine-tuning is about 1 part in $10^5$.  However, if the GUT scale parameters are constrained such that $m_{H_{u,d}}/m_{16} \approx \sqrt{2}$ and $A_0/m_{16}\approx-2$, then the fine-tuning of our theory is reduced to about 1 part in $500$.  This suggests that, in a more fundamental natural theory, the ratio of $m_{16}$ with $m_{H_{u,d}}$ and $A_0$ could be fixed naturally.  In the context of the Bear et.~al.~ argument that one should combine dependent terms into a single independent quantity before evaluating fine-tuning\cite{Baer:2014ica}, we claim that $m_{16}, m_{H_{u,d}}$, and $A_0$ might be dependent quantities in a more fundamental theory.  Hence, one should combine these quantities before calculating fine-tuning.  We discuss one possible partial example in the Appendix.

\section{Conclusion: SUSY on the Edge}
\label{sec:conclusion}
We have analyzed a three family SO(10) SUSY GUT with Yukawa unification for the third family.  The model gives reasonable fits to fermion masses and mixing angles, as well as many other low energy observables;  see Appendix \ref{app:benchmark} with some benchmark points of the global $\chi^2$ analysis. A plot of $\chi^2$ as a function of the parameter $m_{16}$ is given in Figure \ref{fig:chi2_line_plot} or $\chi^2$ contours in the two dimensional plane of $M_{\tilde g}$ vs. $m_{16}$ is given in Figure \ref{fig:contour}.

We performed an analysis with universal gaugino masses and with non-universal gaugino mass with splitting determined by ``mirage mediation" boundary conditions described in Eqn. \ref{eqn:mirage}.  The parameter $\alpha = 0$ for universal gaugino masses and we also take $\alpha = 1.5$ which is consistent with a well-tempered dark matter candidate \cite{Anandakrishnan:2014fia}.  In both cases the model favors $m_{16} \approx 25\ \text{TeV}$.  Nevertheless, due to RG running \cite{Bagger:1999ty}, stops and sbottoms have mass of order $5\ \text{TeV}$, while the first two family scalar masses are of order $m_{16}$.  With $m_{16}$ lying in this mass range, stops in our model do not completely decouple from low energy flavor observables (see Section \ref{ssec:b_physics}).   Best fits are found with a gluino mass less than 2 TeV. Our gluinos decay predominantly into third generation quarks \cite{Anandakrishnan:2014nea}.  Moreover, in a previous analysis \cite{Anandakrishnan:2014nea} we showed that the dominant LHC signature for gluinos in the model is given by $b$-Jets, leptons and missing $E_T$.   Note that, in general, the gauginos are the lightest sparticles.  The CP odd Higgs mass is of order 2 TeV, thus the light Higg couplings are very much Standard Model-like.  In Table \ref{tab:prediction} we present additional predictions.  We give the predictions for the CP violating angle for neutrino oscillations, $\delta$, the branching ratio $BR(\mu \to e \gamma)$ and the electric dipole moment of the electron for two different values of $M_{\tilde g}$ and for $\alpha = 0 \text{and} 1.5$.    In addition, $BR(\mu \to e \gamma)$ and the electric dipole moment of the electron are within reach of future experiments.  Thus this theory is eminently testable!

We evaluated the amount of high scale fine-tuning of our model.  In general we find fine-tuning of order 1 part in $10^5$.  However we note that with particular boundary conditions at the GUT scale (when the ratio of $m_{16}$ to $A_0$ and $m_{H_{u,d}}$ are fixed at $A_0/m_{16}\approx-2$ and $m_{H_{u,d}}/m_{16}\approx\sqrt{2}$) the fine-tuning is reduced to 1 part in $500$. We do not have a fundamental theory that gives these two ratios naturally, Nevertheless, in such a fundamental theory the amount of fine-tuning is reduced considerably.

Finally, with the large value of $m_{16} \sim 25\ \text{TeV}$ we expect the gravitino mass to be at least this large.  Perhaps it is large enough to avoid a cosmological gravitino problem \cite{Weinberg:1982zq}.  In addition, moduli may also be suitably heavy to avoid a cosmological moduli problem \cite{Coughlan:1983ci,Ellis:1986zt,Banks:1993en,deCarlos:1993jw}.  Hence the scalar masses are clearly in an intermediate range, i.e. too heavy to be ``natural" and lighter than ``Split SUSY."   We thus are positioned on the border between these two limiting cases, i.e.  this is ``SUSY on the {\bf Edge}."

\section*{Acknowledgments}

We are indebted to Radovan Derm\'i\v{s}ek for his program and his valuable inputs in using it.  We are also grateful to B. Charles Bryant and Archana Anandakrishnan for discussions.  Z.P.~and S.R.~received partial support for this work from DOE/ DE-SC0011726. We thank the \emph{Ohio Supercomputer Center} for using their resources.

\clearpage
\newpage

\newpage
\appendix
\section{Benchmark Points}
\label{app:benchmark}

\begin{table}[!htbp]\footnotesize
  \caption{Benchmark point with $m_{16} = 25\ \text{TeV}, M_{\tilde g} = 1.202\ \text{TeV}, \alpha=0$:}
  \label{tab:fit25}
  $(1/\alpha_G, M_G, \epsilon_3) = (25.98, 2.55\times10^{16}\ \text{GeV}, -1.30\%)$\\
  $(\lambda, \lambda\epsilon, \sigma, \lambda\tilde\epsilon, \rho, \lambda\epsilon', \lambda\epsilon\xi) = (0.6101, 0.0308, 1.1559, 0.0049, 0.0698, -0.0019, 0.0036)$\\
  $(\phi_\sigma, \phi_{\tilde\epsilon}, \phi_{\rho}, \phi_\xi) = (0.52, 0.58, 3.95, 3.47)\text{rad}$\\
  $(m_{16}, M_{1/2}, A_0, \mu(M_Z)) = (25000, 280, -51380, 1212)\ \text{GeV}$\\
  $((m_{H_d}/m_{16})^2, (m_{H_u}/m_{16})^2, \tan\beta) = (1.86, 1.61, 50.29)$\\
  $(M_{R_1}, M_{R_2}, M_{R_3}) = (9.2, 578.8, 35054.2)\ \times10^9\text{GeV}$

  \begin{center}
    \scalebox{0.8}{
      \begin{tabular}{|l|r|r|r|r|}
        \hline
        Observable & Fit & Exp. & Pull & $\sigma$ \\
        \hline\hline
        $M_Z$ & 91.1876 & 91.1876 & 0.0000 & 0.4541 \\
        $M_W$ & 80.4784 & 80.3850 & 0.2320 & 0.4027 \\
        $1/\alpha_\text{em}$ &  137.2810 & 0.0073 & 0.3569 & 0.6864 \\
        $G_\mu\times10^5$ & 1.1789 & 1.1664 & 1.0598 & 0.0118 \\
        $\alpha_3(M_Z)$ & 0.1192 & 0.1185 & 0.8199 & 0.0008 \\
        \hline
        $M_t$ & 174.0947 & 173.2100 & 0.7171 & 1.2337 \\
        $m_b(m_b)$ & 4.1986 & 4.1800 & 0.5092 & 0.0366 \\
        $m_\tau$ & 1.7772 & 1.7768 & 0.0428 & 0.0089 \\
        \hline
        $M_b-M_c$ & 3.1701 & 3.4500 & 0.8720 & 0.3209 \\
        $m_c(m_c)$ & 1.2509 & 1.2750 & 0.9333 & 0.0258 \\
        $m_s(2\ \text{GeV})$ & 0.0953 & 0.0950 & 0.0609 & 0.0050 \\
        $m_d/m_s(2\ \text{GeV})$  & 0.0702 & 0.0513 & 2.8247 & 0.0067 \\
        $1/Q^2$ & 0.0018 & 0.0019 & 0.4528 & 0.0001 \\
        $M_\mu$ & 0.1056 & 0.1057 & 0.0578 & 0.0005 \\
        $M_e\times10^4$ & 5.1143 & 5.1100 & 0.1674 & 0.0256 \\
        \hline
        $|V_{us}|$ & 0.2245 & 0.2253 & 0.5931 & 0.0014 \\
        $|V_{cb}|$ & 0.0404 & 0.0408 & 0.1670 & 0.0021 \\
        $|V_{ub}|\times10^3$ & 3.1235 & 3.8500 & 0.8446 & 0.8601 \\
        $|V_{td}|\times10^3$ & 8.8463 & 8.4000 & 0.7418 & 0.6016 \\
        $|V_{ts}|$ & 0.0396 & 0.0400 & 0.1508 & 0.0027 \\
        $\sin2\beta$ & 0.6296 & 0.6820 & 2.7214 & 0.0193 \\
        $\epsilon_K$ & 0.0022 & 0.0022 & 0.0022 & 0.0002 \\
        \hline
        $\Delta M_{B_s}/\Delta M_{B_d}$ & 34.8195 & 35.0345 & 0.0308 & 6.9747 \\
        $\Delta M_{B_d}\times10^{13}$ & 3.9946 & 3.3370 & 0.8224 & 0.7996 \\
        \hline
        $m^2_{21}\times10^5$ &   7.5883 & 7.5550 & 0.0621 & 0.5363 \\
        $m^2_{31}\times10^3$ &  2.4649 & 2.4620 & 0.0197 & 0.1455 \\
        $\sin^2\theta_{12}$ & 0.3028 & 0.3070 & 0.1125 & 0.0370 \\
        $\sin^2\theta_{23}$ & 0.6600 & 0.5125 & 1.1300 & 0.1305 \\
        $\sin^2\theta_{13}$ & 0.0162 & 0.0218 & 1.7510 & 0.0032 \\
        \hline
        $M_h$ & 126.2697 & 125.7000 & 0.1882 & 3.0265 \\
        \hline
        $BR(B\to s\gamma)\times10^4$ & 2.7220 & 3.4300 & 0.5419 & 1.3064 \\
        $BR(B_s\to\mu^+\mu^-)\times10^9$ & 2.7213 & 2.8000 & 0.0888 & 0.8867 \\
        $BR(B_d\to\mu^+\mu^-)\times10^{10}$ & 1.0734 & 3.9000 & 1.7509 & 1.6143 \\
        $BR(B\to\tau\nu)\times10^5$ & 6.2223 & 11.4000 & 1.3588 & 3.8104 \\
        \hline
        $BR(B\to K^*\mu^+\mu^-)_{1\leq q^2\leq6\ \text{GeV}^2}\times10^8$ & 4.7860 & 3.4000 & 0.2739 & 5.0610 \\
        $BR(B\to K^*\mu^+\mu^-)_{14.18\leq q^2\leq 16\ \text{GeV}^2}\times10^8$ & 7.5495 & 5.6000 & 0.1356 & 14.3788 \\
        $q_0^2(A_\text{FB}(B\to K^*\mu^+\mu^-))$ & 3.7120 & 4.9000 & 0.9190 & 1.2927 \\
        $F_L(B\to K^*\mu^+\mu^-)_{1\leq q^2\leq6\ \text{GeV}^2}$ & 0.7207 & 0.6500 & 0.2101 & 0.3366 \\
        $F_L(B\to K^*\mu^+\mu^-)_{14.18\leq q^2\leq16\ \text{GeV}^2}$ & 0.3108 & 0.3300 & 0.0726 & 0.2644 \\
        $P_2(B\to K^*\mu^+\mu^-)_{1\leq q^2\leq6\ \text{GeV}^2}$ & 0.0331 & 0.3300 & 2.3939 & 0.1240 \\
        $P_2(B\to K^*\mu^+\mu^-)_{14.18\leq q^2\leq16\ \text{GeV}^2}$ & -0.4336 & -0.5000 & 0.3364 & 0.1974 \\
        $P_4'(B\to K^*\mu^+\mu^-)_{1\leq q^2\leq6\ \text{GeV}^2}$ & 0.5717 & 0.5800 & 0.0208 & 0.3988 \\
        $P_4'(B\to K^*\mu^+\mu^-)_{14.18\leq q^2\leq16\ \text{GeV}^2}$ & 1.2190 & -0.1800 & 1.7066 & 0.8198 \\
        $P_5'(B\to K^*\mu^+\mu^-)_{1\leq q^2\leq6\ \text{GeV}^2}$ & -0.4335 & 0.2100 & 2.2451 & 0.2866 \\
        $P_5'(B\to K^*\mu^+\mu^-)_{14.18\leq q^2\leq16\ \text{GeV}^2}$ & -0.7116 & -0.7900 & 0.1552 & 0.5052 \\
        \hline
        \multicolumn{3}{|l|}{Total $\chi^2$} & 46.7692 & \\
        \hline
      \end{tabular}
    }
  \end{center}
\end{table}

\begin{table}[!htbp]\small
  \caption{Benchmark point with $m_{16} = 25\ \text{TeV}, M_{\tilde g} = 1.187\ \text{TeV}, \alpha=1.5$:}
  \label{tab:fit25A}
  $(1/\alpha_G, M_G, \epsilon_3) = (25.95, 2.60\times10^{16}\ \text{GeV}, -1.50\%)$\\
  $(\lambda, \lambda\epsilon, \sigma, \lambda\tilde\epsilon, \rho, \lambda\epsilon', \lambda\epsilon\xi) = (0.6100, 0.0310, 1.1459, 0.0049, 0.0708, -0.0019, 0.0037)$\\
  $(\phi_\sigma, \phi_{\tilde\epsilon}, \phi_{\rho}, \phi_\xi) = (0.53, 0.57, 3.94, 3.49)\text{rad}$\\
  $(m_{16}, M_{1/2}, A_0, \mu(M_Z)) = (25000, 520, -51157, 1236)\ \text{GeV}$\\
  $((m_{H_d}/m_{16})^2, (m_{H_u}/m_{16})^2, \tan\beta) = (1.85, 1.61, 50.19)$\\
  $(M_{R_1}, M_{R_2}, M_{R_3}) = (9.1, 567.0, 32370.5)\ \times10^9\text{GeV}$

  \begin{center}
    \scalebox{0.8}{
      \begin{tabular}{|l|r|r|r|r|}
        \hline
        Observable & Fit & Exp. & Pull & $\sigma$ \\
        \hline\hline
        $M_Z$ & 91.1876 & 91.1876 & 0.0000 & 0.4540 \\
        $M_W$ & 80.5197 & 80.3850 & 0.3344 & 0.4029 \\
        $1/\alpha_\text{em}$ &  137.1416 & 0.0073 & 0.1540 & 0.6857 \\
        $G_\mu\times10^5$ & 1.1829 & 1.1664 & 1.3978 & 0.0118 \\
        $\alpha_3(M_Z)$ & 0.1189 & 0.1185 & 0.4798 & 0.0008 \\
        \hline
        $M_t$ & 173.8449 & 173.2100 & 0.5150 & 1.2328 \\
        $m_b(m_b)$ & 4.2023 & 4.1800 & 0.6094 & 0.0366 \\
        $m_\tau$ & 1.7772 & 1.7768 & 0.0450 & 0.0089 \\
        \hline
        $M_b-M_c$ & 3.1680 & 3.4500 & 0.8791 & 0.3207 \\
        $m_c(m_c)$ & 1.2570 & 1.2750 & 0.6979 & 0.0258 \\
        $m_s(2\ \text{GeV})$ & 0.0947 & 0.0950 & 0.0671 & 0.0050 \\
        $m_d/m_s(2\ \text{GeV})$  & 0.0700 & 0.0513 & 2.7901 & 0.0067 \\
        $1/Q^2$ & 0.0018 & 0.0019 & 0.5027 & 0.0001 \\
        $M_\mu$ & 0.1056 & 0.1057 & 0.1457 & 0.0005 \\
        $M_e\times10^4$ & 5.1145 & 5.1100 & 0.1775 & 0.0256 \\
        \hline
        $|V_{us}|$ & 0.2244 & 0.2253 & 0.6440 & 0.0014 \\
        $|V_{cb}|$ & 0.0407 & 0.0408 & 0.0584 & 0.0021 \\
        $|V_{ub}|\times10^3$ & 3.1307 & 3.8500 & 0.8363 & 0.8601 \\
        $|V_{td}|\times10^3$ & 8.8596 & 8.4000 & 0.7639 & 0.6016 \\
        $|V_{ts}|$ & 0.0398 & 0.0400 & 0.0652 & 0.0027 \\
        $\sin2\beta$ & 0.6285 & 0.6820 & 2.7790 & 0.0193 \\
        $\epsilon_K$ & 0.0023 & 0.0022 & 0.1149 & 0.0002 \\
        \hline
        $\Delta M_{B_s}/\Delta M_{B_d}$ & 35.5946 & 35.0345 & 0.0786 & 7.1295 \\
        $\Delta M_{B_d}\times10^{13}$ & 3.9756 & 3.3370 & 0.8025 & 0.7958 \\
        \hline
        $m^2_{21}\times10^5$ &   7.6111 & 7.5550 & 0.1046 & 0.5364 \\
        $m^2_{31}\times10^3$ &  2.4657 & 2.4620 & 0.0255 & 0.1455 \\
        $\sin^2\theta_{12}$ & 0.3134 & 0.3070 & 0.1724 & 0.0370 \\
        $\sin^2\theta_{23}$ & 0.6319 & 0.5125 & 0.9146 & 0.1305 \\
        $\sin^2\theta_{13}$ & 0.0153 & 0.0218 & 2.0337 & 0.0032 \\
        \hline
        $M_h$ & 124.5455 & 125.7000 & 0.3814 & 3.0265 \\
        \hline
        $BR(B\to s\gamma)\times10^4$ & 2.7270 & 3.4300 & 0.5372 & 1.3087 \\
        $BR(B_s\to\mu^+\mu^-)\times10^9$ & 2.5215 & 2.8000 & 0.3228 & 0.8627 \\
        $BR(B_d\to\mu^+\mu^-)\times10^{10}$ & 1.0192 & 3.9000 & 1.7861 & 1.6129 \\
        $BR(B\to\tau\nu)\times10^5$ & 6.2272 & 11.4000 & 1.3568 & 3.8124 \\
        \hline
        $BR(B\to K^*\mu^+\mu^-)_{1\leq q^2\leq6\ \text{GeV}^2}\times10^8$ & 4.8580 & 3.4000 & 0.2839 & 5.1361 \\
        $BR(B\to K^*\mu^+\mu^-)_{14.18\leq q^2\leq 16\ \text{GeV}^2}\times10^8$ & 7.6648 & 5.6000 & 0.1415 & 14.5975 \\
        $q_0^2(A_\text{FB}(B\to K^*\mu^+\mu^-))$ & 3.7150 & 4.9000 & 0.9163 & 1.2933 \\
        $F_L(B\to K^*\mu^+\mu^-)_{1\leq q^2\leq6\ \text{GeV}^2}$ & 0.7208 & 0.6500 & 0.2103 & 0.3366 \\
        $F_L(B\to K^*\mu^+\mu^-)_{14.18\leq q^2\leq16\ \text{GeV}^2}$ & 0.3108 & 0.3300 & 0.0726 & 0.2644 \\
        $P_2(B\to K^*\mu^+\mu^-)_{1\leq q^2\leq6\ \text{GeV}^2}$ & 0.0335 & 0.3300 & 2.3879 & 0.1242 \\
        $P_2(B\to K^*\mu^+\mu^-)_{14.18\leq q^2\leq16\ \text{GeV}^2}$ & -0.4336 & -0.5000 & 0.3364 & 0.1974 \\
        $P_4'(B\to K^*\mu^+\mu^-)_{1\leq q^2\leq6\ \text{GeV}^2}$ & 0.5697 & 0.5800 & 0.0258 & 0.3985 \\
        $P_4'(B\to K^*\mu^+\mu^-)_{14.18\leq q^2\leq16\ \text{GeV}^2}$ & 1.2190 & -0.1800 & 1.7066 & 0.8198 \\
        $P_5'(B\to K^*\mu^+\mu^-)_{1\leq q^2\leq6\ \text{GeV}^2}$ & -0.4334 & 0.2100 & 2.2450 & 0.2866 \\
        $P_5'(B\to K^*\mu^+\mu^-)_{14.18\leq q^2\leq16\ \text{GeV}^2}$ & -0.7117 & -0.7900 & 0.1550 & 0.5052 \\
        \hline
        \multicolumn{3}{|l|}{Total $\chi^2$} & 47.7692 & \\
        \hline
      \end{tabular}
    }
  \end{center}
\end{table}

\begin{table}[!htbp]\small
  \caption{Benchmark point with $m_{16} = 25\ \text{TeV}, M_{\tilde g} = 1.613\ \text{TeV}, \alpha=0$:}
  \label{tab:fit25A}
  $(1/\alpha_G, M_G, \epsilon_3) = (26.22, 2.32\times10^{16}\ \text{GeV}, -0.65\%)$\\
  $(\lambda, \lambda\epsilon, \sigma, \lambda\tilde\epsilon, \rho, \lambda\epsilon', \lambda\epsilon\xi) = (0.6096, 0.0311, 1.1398, 0.0049, 0.0710, -0.0019, 0.0038)$\\
  $(\phi_\sigma, \phi_{\tilde\epsilon}, \phi_{\rho}, \phi_\xi) = (0.53, 0.56, 3.95, 3.49)\text{rad}$\\
  $(m_{16}, M_{1/2}, A_0, \mu(M_Z)) = (25000, 450, -51341, 1226)\ \text{GeV}$\\
  $((m_{H_d}/m_{16})^2, (m_{H_u}/m_{16})^2, \tan\beta) = (1.86, 1.61, 50.30)$\\
  $(M_{R_1}, M_{R_2}, M_{R_3}) = (9.1, 572.4, 32277.4)\ \times10^9\text{GeV}$

  \begin{center}
    \scalebox{0.8}{
      \begin{tabular}{|l|r|r|r|r|}
        \hline
        Observable & Fit & Exp. & Pull & $\sigma$ \\
        \hline\hline
        $M_Z$ & 91.1876 & 91.1876 & 0.0000 & 0.4535 \\
        $M_W$ & 80.4507 & 80.3850 & 0.1633 & 0.4025 \\
        $1/\alpha_\text{em}$ &  137.7125 & 0.0073 & 0.9825 & 0.6886 \\
        $G_\mu\times10^5$ & 1.1732 & 1.1664 & 0.5798 & 0.0117 \\
        $\alpha_3(M_Z)$ & 0.1188 & 0.1185 & 0.4140 & 0.0008 \\
        \hline
        $M_t$ & 174.1882 & 173.2100 & 0.7927 & 1.2340 \\
        $m_b(m_b)$ & 4.1954 & 4.1800 & 0.4220 & 0.0366 \\
        $m_\tau$ & 1.7781 & 1.7768 & 0.1417 & 0.0089 \\
        \hline
        $M_b-M_c$ & 3.1568 & 3.4500 & 0.9175 & 0.3196 \\
        $m_c(m_c)$ & 1.2595 & 1.2750 & 0.5993 & 0.0258 \\
        $m_s(2\ \text{GeV})$ & 0.0939 & 0.0950 & 0.2147 & 0.0050 \\
        $m_d/m_s(2\ \text{GeV})$  & 0.0701 & 0.0513 & 2.8052 & 0.0067 \\
        $1/Q^2$ & 0.0018 & 0.0019 & 0.5139 & 0.0001 \\
        $M_\mu$ & 0.1056 & 0.1057 & 0.1818 & 0.0005 \\
        $M_e\times10^4$ & 5.1145 & 5.1100 & 0.1749 & 0.0256 \\
        \hline
        $|V_{us}|$ & 0.2244 & 0.2253 & 0.6763 & 0.0014 \\
        $|V_{cb}|$ & 0.0404 & 0.0408 & 0.1729 & 0.0021 \\
        $|V_{ub}|\times10^3$ & 3.1033 & 3.8500 & 0.8681 & 0.8601 \\
        $|V_{td}|\times10^3$ & 8.8101 & 8.4000 & 0.6817 & 0.6016 \\
        $|V_{ts}|$ & 0.0396 & 0.0400 & 0.1531 & 0.0027 \\
        $\sin2\beta$ & 0.6270 & 0.6820 & 2.8562 & 0.0193 \\
        $\epsilon_K$ & 0.0022 & 0.0022 & 0.2052 & 0.0002 \\
        \hline
        $\Delta M_{B_s}/\Delta M_{B_d}$ & 35.3739 & 35.0345 & 0.0479 & 7.0854 \\
        $\Delta M_{B_d}\times10^{13}$ & 3.9433 & 3.3370 & 0.7681 & 0.7894 \\
        \hline
        $m^2_{21}\times10^5$ &   7.6562 & 7.5550 & 0.1886 & 0.5364 \\
        $m^2_{31}\times10^3$ &  2.4631 & 2.4620 & 0.0077 & 0.1455 \\
        $\sin^2\theta_{12}$ & 0.3170 & 0.3070 & 0.2689 & 0.0370 \\
        $\sin^2\theta_{23}$ & 0.6264 & 0.5125 & 0.8722 & 0.1305 \\
        $\sin^2\theta_{13}$ & 0.0149 & 0.0218 & 2.1658 & 0.0032 \\
        \hline
        $M_h$ & 124.5054 & 125.7000 & 0.3947 & 3.0265 \\
        \hline
        $BR(B\to s\gamma)\times10^4$ & 2.6840 & 3.4300 & 0.5789 & 1.2887 \\
        $BR(B_s\to\mu^+\mu^-)\times10^9$ & 3.0247 & 2.8000 & 0.2429 & 0.9252 \\
        $BR(B_d\to\mu^+\mu^-)\times10^{10}$ & 1.1022 & 3.9000 & 1.7323 & 1.6151 \\
        $BR(B\to\tau\nu)\times10^5$ & 6.1884 & 11.4000 & 1.3727 & 3.7966 \\
        \hline
        $BR(B\to K^*\mu^+\mu^-)_{1\leq q^2\leq6\ \text{GeV}^2}\times10^8$ & 4.7640 & 3.4000 & 0.2707 & 5.0381 \\
        $BR(B\to K^*\mu^+\mu^-)_{14.18\leq q^2\leq 16\ \text{GeV}^2}\times10^8$ & 7.5110 & 5.6000 & 0.1336 & 14.3059 \\
        $q_0^2(A_\text{FB}(B\to K^*\mu^+\mu^-))$ & 3.6690 & 4.9000 & 0.9579 & 1.2850 \\
        $F_L(B\to K^*\mu^+\mu^-)_{1\leq q^2\leq6\ \text{GeV}^2}$ & 0.7225 & 0.6500 & 0.2149 & 0.3374 \\
        $F_L(B\to K^*\mu^+\mu^-)_{14.18\leq q^2\leq16\ \text{GeV}^2}$ & 0.3108 & 0.3300 & 0.0726 & 0.2644 \\
        $P_2(B\to K^*\mu^+\mu^-)_{1\leq q^2\leq6\ \text{GeV}^2}$ & 0.0228 & 0.3300 & 2.5196 & 0.1219 \\
        $P_2(B\to K^*\mu^+\mu^-)_{14.18\leq q^2\leq16\ \text{GeV}^2}$ & -0.4336 & -0.5000 & 0.3364 & 0.1974 \\
        $P_4'(B\to K^*\mu^+\mu^-)_{1\leq q^2\leq6\ \text{GeV}^2}$ & 0.5820 & 0.5800 & 0.0050 & 0.4001 \\
        $P_4'(B\to K^*\mu^+\mu^-)_{14.18\leq q^2\leq16\ \text{GeV}^2}$ & 1.2190 & -0.1800 & 1.7066 & 0.8198 \\
        $P_5'(B\to K^*\mu^+\mu^-)_{1\leq q^2\leq6\ \text{GeV}^2}$ & -0.4455 & 0.2100 & 2.2578 & 0.2903 \\
        $P_5'(B\to K^*\mu^+\mu^-)_{14.18\leq q^2\leq16\ \text{GeV}^2}$ & -0.7116 & -0.7900 & 0.1552 & 0.5052 \\
        \hline
        \multicolumn{3}{|l|}{Total $\chi^2$} & 48.8413 & \\
        \hline
      \end{tabular}
    }
  \end{center}
\end{table}

\begin{table}[!htbp]\small
  \caption{Benchmark point with $m_{16} = 25\ \text{TeV}, M_{\tilde g} = 1.690\ \text{TeV}, \alpha=1.5$:}
  \label{tab:fit25A}
  $(1/\alpha_G, M_G, \epsilon_3) = (26.38, 2.09\times10^{16}\ \text{GeV}, 0.02\%)$\\
  $(\lambda, \lambda\epsilon, \sigma, \lambda\tilde\epsilon, \rho, \lambda\epsilon', \lambda\epsilon\xi) = (0.6096, 0.0311, 1.1384, 0.0049, 0.0708, -0.0019, 0.0037)$\\
  $(\phi_\sigma, \phi_{\tilde\epsilon}, \phi_{\rho}, \phi_\xi) = (0.52, 0.56, 3.96, 3.49)\text{rad}$\\
  $(m_{16}, M_{1/2}, A_0, \mu(M_Z)) = (25000, 900, -50846, 1529)\ \text{GeV}$\\
  $((m_{H_d}/m_{16})^2, (m_{H_u}/m_{16})^2, \tan\beta) = (1.86, 1.60, 50.31)$\\
  $(M_{R_1}, M_{R_2}, M_{R_3}) = (9.1, 579.0, 32367.3)\ \times10^9\text{GeV}$

  \begin{center}
    \scalebox{0.8}{
      \begin{tabular}{|l|r|r|r|r|}
        \hline
        Observable & Fit & Exp. & Pull & $\sigma$ \\
        \hline\hline
        $M_Z$ & 91.1876 & 91.1876 & 0.0000 & 0.4540 \\
        $M_W$ & 80.4655 & 80.3850 & 0.2000 & 0.4026 \\
        $1/\alpha_\text{em}$ &  137.7323 & 0.0073 & 1.0111 & 0.6887 \\
        $G_\mu\times10^5$ & 1.1740 & 1.1664 & 0.6469 & 0.0117 \\
        $\alpha_3(M_Z)$ & 0.1188 & 0.1185 & 0.2979 & 0.0008 \\
        \hline
        $M_t$ & 174.3427 & 173.2100 & 0.9175 & 1.2345 \\
        $m_b(m_b)$ & 4.2001 & 4.1800 & 0.5479 & 0.0366 \\
        $m_\tau$ & 1.7774 & 1.7768 & 0.0644 & 0.0089 \\
        \hline
        $M_b-M_c$ & 3.1659 & 3.4500 & 0.8863 & 0.3205 \\
        $m_c(m_c)$ & 1.2574 & 1.2750 & 0.6825 & 0.0258 \\
        $m_s(2\ \text{GeV})$ & 0.0936 & 0.0950 & 0.2741 & 0.0050 \\
        $m_d/m_s(2\ \text{GeV})$  & 0.0701 & 0.0513 & 2.8082 & 0.0067 \\
        $1/Q^2$ & 0.0018 & 0.0019 & 0.5170 & 0.0001 \\
        $M_\mu$ & 0.1056 & 0.1057 & 0.1571 & 0.0005 \\
        $M_e\times10^4$ & 5.1139 & 5.1100 & 0.1545 & 0.0256 \\
        \hline
        $|V_{us}|$ & 0.2244 & 0.2253 & 0.6688 & 0.0014 \\
        $|V_{cb}|$ & 0.0400 & 0.0408 & 0.3609 & 0.0021 \\
        $|V_{ub}|\times10^3$ & 3.0662 & 3.8500 & 0.9113 & 0.8601 \\
        $|V_{td}|\times10^3$ & 8.7156 & 8.4000 & 0.5247 & 0.6016 \\
        $|V_{ts}|$ & 0.0392 & 0.0400 & 0.2960 & 0.0027 \\
        $\sin2\beta$ & 0.6259 & 0.6820 & 2.9122 & 0.0193 \\
        $\epsilon_K$ & 0.0022 & 0.0022 & 0.0834 & 0.0002 \\
        \hline
        $\Delta M_{B_s}/\Delta M_{B_d}$ & 34.7964 & 35.0345 & 0.0342 & 6.9701 \\
        $\Delta M_{B_d}\times10^{13}$ & 3.8958 & 3.3370 & 0.7165 & 0.7799 \\
        \hline
        $m^2_{21}\times10^5$ &   7.6614 & 7.5550 & 0.1984 & 0.5364 \\
        $m^2_{31}\times10^3$ &  2.4606 & 2.4620 & 0.0094 & 0.1455 \\
        $\sin^2\theta_{12}$ & 0.3197 & 0.3070 & 0.3423 & 0.0370 \\
        $\sin^2\theta_{23}$ & 0.6197 & 0.5125 & 0.8210 & 0.1305 \\
        $\sin^2\theta_{13}$ & 0.0146 & 0.0218 & 2.2520 & 0.0032 \\
        \hline
        $M_h$ & 122.0502 & 125.7000 & 1.2059 & 3.0265 \\
        \hline
        $BR(B\to s\gamma)\times10^4$ & 2.6310 & 3.4300 & 0.6321 & 1.2640 \\
        $BR(B_s\to\mu^+\mu^-)\times10^9$ & 3.5145 & 2.8000 & 0.7203 & 0.9920 \\
        $BR(B_d\to\mu^+\mu^-)\times10^{10}$ & 1.0522 & 3.9000 & 1.7647 & 1.6138 \\
        $BR(B\to\tau\nu)\times10^5$ & 6.1009 & 11.4000 & 1.4090 & 3.7610 \\
        \hline
        $BR(B\to K^*\mu^+\mu^-)_{1\leq q^2\leq6\ \text{GeV}^2}\times10^8$ & 4.6780 & 3.4000 & 0.2583 & 4.9484 \\
        $BR(B\to K^*\mu^+\mu^-)_{14.18\leq q^2\leq 16\ \text{GeV}^2}\times10^8$ & 7.4066 & 5.6000 & 0.1281 & 14.1080 \\
        $q_0^2(A_\text{FB}(B\to K^*\mu^+\mu^-))$ & 3.6290 & 4.9000 & 0.9946 & 1.2779 \\
        $F_L(B\to K^*\mu^+\mu^-)_{1\leq q^2\leq6\ \text{GeV}^2}$ & 0.7240 & 0.6500 & 0.2189 & 0.3380 \\
        $F_L(B\to K^*\mu^+\mu^-)_{14.18\leq q^2\leq16\ \text{GeV}^2}$ & 0.3108 & 0.3300 & 0.0726 & 0.2644 \\
        $P_2(B\to K^*\mu^+\mu^-)_{1\leq q^2\leq6\ \text{GeV}^2}$ & 0.0132 & 0.3300 & 2.6254 & 0.1207 \\
        $P_2(B\to K^*\mu^+\mu^-)_{14.18\leq q^2\leq16\ \text{GeV}^2}$ & -0.4337 & -0.5000 & 0.3358 & 0.1975 \\
        $P_4'(B\to K^*\mu^+\mu^-)_{1\leq q^2\leq6\ \text{GeV}^2}$ & 0.5918 & 0.5800 & 0.0294 & 0.4014 \\
        $P_4'(B\to K^*\mu^+\mu^-)_{14.18\leq q^2\leq16\ \text{GeV}^2}$ & 1.2190 & -0.1800 & 1.7066 & 0.8198 \\
        $P_5'(B\to K^*\mu^+\mu^-)_{1\leq q^2\leq6\ \text{GeV}^2}$ & -0.4562 & 0.2100 & 2.2685 & 0.2937 \\
        $P_5'(B\to K^*\mu^+\mu^-)_{14.18\leq q^2\leq16\ \text{GeV}^2}$ & -0.7117 & -0.7900 & 0.1550 & 0.5052 \\
        \hline
        \multicolumn{3}{|l|}{Total $\chi^2$} & 52.6056 & \\
        \hline
      \end{tabular}
    }
  \end{center}
\end{table}

\section{Natural SUSY Breaking}
\label{app:string_natural}

Consider Heterotic orbifold models with dilaton/moduli SUSY breaking as discussed in \cite{Brignole:1997dp}.  Following Eqn. (60, 61) in this paper we have the scalar masses for sparticle $\alpha$, the trilinear couplings and the gaugino masses are given by
\begin{align}
  m^2_{\alpha} &= m_{3/2}^2(1 + 3 C^2 \cos^2\theta \vec{n}_\alpha\cdot\vec{\Theta}^2)+V_0
  \\A_0 &= - \sqrt{3} C m_{3/2}(\sin\theta e^{i\gamma s}+\cos\theta \sum_{i=1}^6e^{-i\gamma_i}\Theta_i[1+n_\alpha^{T_i}+n_\beta^{T_i}+n_\gamma^{T_i}-(T_i+T_i^*)\partial_i\log Y_{161016}]) \,,
  \\ M_a &  = \sqrt{3} m_{3/2} \sin\theta e^{-i\gamma s} \,,
\end{align}
where
\begin{align}
  C^2 = 1 + \frac{V_0}{3m_{10}^2} \,,
\end{align}
$V_0$ is the tree-level cosmological constant that we will set to be zero, i.e. $C = 1$ and $\theta$ determines the amount of SUSY breaking of the dilaton sector versus the moduli sector.  $\cos\theta=0$ means all the SUSY breaking is due to the dilaton.  $\vec{n}_\alpha$ are the modular weights of matter fields and $\vec{\Theta}$ gives the probability for the SUSY breaking contribution of each modulus, such that $\sum_{i=1}^6 \Theta_i^2 = 1$.
We then let $\theta = 0$ such that at tree level gauginos are massless.   They would then obtain one loop suppressed masses due to moduli SUSY breaking.  We also assume that the Yukawa couplings, $Y_{\alpha \beta \gamma}$ are independent of the moduli.   As a particular example,  consider a $\mathbb{Z}_2 \otimes \mathbb{Z}_6^\prime$ orbifold with twist vectors given by $(1/2,1/2,0), \ (1/6,2/3,1/6)$ in the three two torii \cite{Ibanez:1992hc}.  There are three Kahler moduli in this example.  Then we have
\begin{align}
  m^2_{\alpha} &= m_{3/2}^2(1 + 3  (n_\alpha^{T_1} \Theta_1^2 + n_\alpha^{T_3} \Theta_3^2))
\end{align}
where we assumed that $\Theta_2 = 0$.

We now assume that the Higgs $10$-plet comes from the bulk on the second two torus.  Thus $n_{10}^{T_1} = n_{10}^{T_3} = 0, \ n_{10}^{T_2} = -1$.   Hence, $m_{10} = m_{3/2}$ and therefore, $m_{16}^2 = m_{10}^2 ( 1 + 3 (n_\alpha^{T_1} \Theta_1^2 + n_\alpha^{T_3} \Theta_3^2))$. If the $16$-plet lives in the fifth twisted sector with modular weights, $n_{16}^{T_1} = n_{16}^{T_3} = -1/6, \ n_{16}^{T_2} = -2/3$, we have $m_{16}^2 = \frac{1}{2} m_{10}^2$ or
$m_{10} = \sqrt{2} m_{16}$.  This is our first constraint, that $m_{10} \equiv (m_{H_u} + m_{H_d})/2 \approx \sqrt{2} m_{16}$.

We then have $A_{(16 \ 16 \ 10)} =  - \sqrt{6} m_{16}(e^{-i\gamma_1}\Theta_1[1+ 2 n_{16}^{T_1} + n_{10}^{T_1}] + e^{-i\gamma_3}\Theta_3[1+ 2 n_{16}^{T_3} + n_{10}^{T_3}])$ and we need $A_{(16 \ 16 \ 10)} = A_0 = -2 m_{16}$ where the last equality is the boundary condition at the sweet spot.  This can be solved with $e^{-i\gamma_1}\Theta_1 = \frac{1}{\sqrt{2}} e^{-i\gamma}, \ e^{-i\gamma_3}\Theta_3 = \frac{1}{\sqrt{2}} e^{+i\gamma}$ and $\gamma = \pi/6$.   Of course, it would require the dynamics of stabilizing moduli and SUSY breaking to fix these particular values of $\Theta_i$.

\clearpage
\newpage

\bibliographystyle{utphys}
\bibliography{bibliography}

\end{document}